\documentstyle[12pt,psfig]{article}
\begin{document}

\vskip 1truecm
\rightline{Preprint DAMTP 96-77, PUPT-1640}
\rightline{ e-Print Archive: hep-ph/9608350}
\vspace{0.6in}
\centerline{\Large  Classical Field Dynamics of the 
		Electroweak Phase Transition}
\vspace{0.6in}
\centerline{\Large Guy D. Moore\footnote{e-mail:
guymoore@puhep1.princeton.edu }
}
\medskip

\centerline{\it Princeton University}
\centerline{\it Joseph Henry Laboratories, PO Box 708}
\centerline{\it Princeton, NJ 08544, USA}

\vspace{0.6in}

\centerline{\Large Neil Turok \footnote{e-mail: 
N.G.Turok@damtp.cam.ac.uk }
}
\medskip

\centerline{\it DAMTP, Silver Street}
\centerline{\it Cambridge, CB3 9EW, UK}

\vspace{0.6in}

\centerline{\bf Abstract}

\smallskip

We investigate the thermodynamics and dynamics of the electroweak phase
transition by modelling the infrared physics with classical Yang-Mills
Higgs theory.  We discuss the accuracy of this approach and conclude
that, for quantities whose determination is dominated by the infrared,
the classical method should be correct up to parametrically suppressed
(ie $O(\alpha)$) corrections.  For a Higgs self-coupling which at tree
level corresponds to $m_H \simeq 50$GeV, we determine the jump in the
order parameter to be $\delta \phi = 1.5 gT$, the surface tension 
to be $\sigma = 0.07 g^4 T^3$, and the friction coefficient on the moving
bubble wall due to infrared bosons 
to be $\eta \equiv P/v_w = 0.03 \pm .004 g^6 T^4$.  We also investigate 
the response of Chern-Simons number to a spatially uniform 
chemical potential and
find that it falls off a short distance inside the bubble wall, both
in equilibrium and below the equilibrium temperature. 

\vspace{0.1in}


\vspace{0.1in}


\vspace{0.4in}

\pagebreak

\section{Introduction}

Since Sakharov proposed that the universe's baryon number
could have been generated dynamically early in the Big Bang \cite{Sakharov},
many different particle physics scenarios have been proposed to realize
this mechanism. One of the most interesting is based on the 
observation made
20 years ago by t'Hooft
that baryon number is violated already in the minimal standard 
model \cite{tHooft}; and by the later realization that the
violation is very
efficient at temperatures above the electroweak phase transition 
temperature, where the Higgs condensate is absent \cite{KRS,McLerran}.
It was subsequently shown that electroweak baryogenesis 
could actually occur in minimal extensions of the standard model
involving one or more additional Higgs doublets, 
and that the resulting asymmetry was naturally
of the order of magnitude required in cosmology \cite{TurokZadrozny},
\cite{MSTV}. Other, potentially more efficient baryogenesis mechanisms 
were proposed \cite{CKN,DHSS}, both in these theories and
in other extensions
of the standard model. What is most exciting about these scenarios is
that they link the origin of
matter in the universe to physics beyond the standard model 
that could be probed by the 
next generation of Higgs-seeking accelerators, and other 
experiments looking for novel sources of C and CP violation.

Because of these developments there has been quite intense interest
in the electroweak phase transition.  But the quest to understand
baryogenesis at the electroweak scale has been bedeviled by difficulties.
It is now believed that we know what basic issues are involved.
As temperature drops, the Higgs field develops a condensate abruptly
in a first order phase transition.  This proceeds by the nucleation and
subsequent growth of bubbles of the low temperature, broken symmetry phase.
Outside the bubbles, baryon number is violated, but inside it is conserved.
The plasma departs from equilibrium on and very near the boundaries of
these bubbles, and if there is some CP violation, for instance in the
Higgs sector, then all necessary conditions for the creation of a 
baryon number excess exist; and it would be preserved to this day because
the rate of baryon number violation is very low inside the bubbles,
if the phase transition is suitably strong.  It is also strongly believed
that the communication of the C and CP violating physics to the infrared 
gauge-Higgs fields, which are responsible for the baryon number violation,
should be conducted by the fermions,
and that the transport of the
fermions in the presence of the wall may be relevant.
But turning this understanding into predictions requires an
accurate description 
of the dynamics of the phase transition, which has so far been 
poorly developed.  The dynamics, in turn,
cannot be well understood until the thermodynamics are under 
control; and all aspects require a complete knowledge of the underlying
electroweak physics, which of course is still lacking.

In the coming years we will hopefully learn the required information 
about electroweak physics, for instance the exact nature of the Higgs
mechanism, the existence or absence of low energy supersymmetry, and
details of any non-CKM CP violation.  But our ignorance of these 
details does not prevent work on the thermodynamics and dynamics of
the electroweak phase transition from proceeding.  The tools developed
to investigate the simplest case, the minimal standard model, should
be straightforward to extend to more complicated models; in fact the
extension may often be analytically tractable, as seems 
to be the case generically for the thermodynamics \cite{KLRS,MSSMguys}.

One tool which has proven less useful than hoped is 
perturbation theory.  While one loop perturbation theory can determine the
phase transition temperature with $O(\alpha)$ accuracy \cite{DolanJackiw},
it is less reliable for the details of the phase transition, such as the
jump in the order parameter, the latent heat, the surface tension, and
so forth.  This became clear when Arnold and Espinosa computed the 
effective potential to two loops \cite{Espinosa,Fodor}.  At one
loop a term of form $-g^3 \phi^3 T/\pi$ arises which generates a Higgs
vev squared of $\phi^2 \sim g^6T^2/\lambda^2 \pi^2$; and at two loops a new
term of form $-g^4 \phi^2 T^2 \ln(\phi/T) /\pi^2$ arises, which alone
would generate $\phi^2 \sim g^4 T^2/\lambda \pi^2$.  As far as the strength
of the phase transition is concerned, perturbation theory is at best an
expansion in $\lambda/g^2$, or $(m_H/m_W)^2$.  

Because $\lambda$ receives large radiative corrections, it is difficult
to make the ratio $\lambda/g^2$ very small; we also know that, if
the minimal standard model is correct, the ratio cannot be small,
because direct searches for the Higgs boson have ruled it out below
a mass of $60$GeV.  It therefore seems reasonable to take $\lambda \sim g^2$
parametrically.  This is also natural from the point of view of unit
analysis, if we allow $\hbar$ to be dimensionful.  To do otherwise would
require a rearrangement of perturbation theory away from the usual
loopwise expansion, which can be understood in the vacuum theory as an 
expansion in powers of $\hbar$.  We see from the above discussion,
though, that at the phase transition temperature, for quantities such
as the jump in the order parameter, the perturbation series is not
an expansion in $\hbar$, which helps to explain its poor convergence.

Farakos et. al. have elucidated the reason for this, by
showing how the infrared thermodynamics of the minimal standard model at
finite temperature is very well described by a ``dimensionally reduced''
3 dimensional field theory \cite{FKRS1}.  this analysis has allowed an
efficient nonperturbative investigation of the phase transition which is 
now bearing fruit \cite{FKRS,KLRS,KLRSresults}.

An at first sight unrelated development is the idea that the rate of
anomalous baryon number violation in the high temperature phase 
of the classical theory is computable numerically, and that the rate in
the classical theory should be the same, up to $O(\alpha)$ corrections,
as the rate in the quantum theory \cite{GrigRub,Amb1,Ambjornetal}.
This has allowed the first quantitative measurement of a dynamical, 
infrared property of finite temperature Yang-Mills Higgs theory above the
phase transition temperature.  The validity of this idea is fortified
by the observation that the thermodynamics of the classical theory
coincides with the thermodynamics of the quantum theory in the
approximation of dimensional reduction \cite{AmbKras}.

In fact we believe that the success of the dimensional reduction technique,
and the nonperturbative nature of the infrared quantum theory, arise
precisely because the infrared bosonic modes attain large occupation 
numbers which cause the theory to mimic a classical theory.  We will
argue in Section \ref{defendclassical} that this mimicry extends to the
dynamics, so that all infrared dominated dynamical properties of the
theory which possess cutoff independent limits in the classical theory
will take the same value, up to $O(\alpha)$ corrections, in the 
quantum theory.  We then review an evolution algorithm for the lattice
cut off, classical field theory in Section \ref{Numerics}.  Since the
classical theory has the same thermodynamics as the quantum theory in the
dimensional reduction approximation, this evolution algorithm is also
a microcanonical Monte-Carlo algorithm for the dimensionally reduced
theory; so we develop and apply the tools for using it to investigate
the thermodynamics of the phase transition in Section \ref{thermodynamics}.
Then we turn to dynamics; in Section \ref{friction} we use the classical
technique to compute the contribution of infrared bosons to the friction
felt by a moving bubble wall as it sweeps through the plasma, in a
near equilibrium approximation.  In Section \ref{NCSsection}
we investigate baryon number violation in the presence of an out of 
equilibrium bubble wall.  Section \ref{conclusion} concludes.
There are also two appendicies.  Appendix A presents the tools for
the numerical study of Chern-Simons number ($N_{CS}$) motion, and investigates
baryon number violation in each phase; in particular it presents evidence
that the observed rate of baryon number violation in the broken phase
arises from ultraviolet lattice artefacts.  Appendix B presents the details
of how to extract the surface tension of the bubble wall.

We make no serious attempt to extrapolate to the continuum limit either
for the thermodynamical or dynamical properties, so the results presented
here should be considered rough and preliminary; the emphasis is
on development of techniques.

\section{Discussion of the classical approximation}
\label{defendclassical}

As discussed in the introduction, we will investigate the use of 
classical Yang-Mills Higgs theory as a surrogate for the Standard Model.
In important respects the classical theory does not resemble the quantum
theory at all (for instance, it has an infinite, or cutoff dependent,
heat capacity); it should only be used for those dynamical 
and thermodynamical properties
which are dominated by infrared physics, where it should
give results which reproduce the parametrically leading
term in the full quantum theory.  Hence, for instance, the classical
theory has been used to investigate baryon number violation in the
symmetric electroweak phase, a phenomenon thought to be infrared
dominated \cite{GrigRub,Amb1,Ambjornetal,AmbKras,Moore1,Moore2,newguys}. 
It has also been used, in a slight disguise, to investigate the
thermodynamics of the phase transition.

To see the latter point, consider the dimensional reduction program of
Farakos et al \cite{early,FKRS,KLRS,KLRSresults}, a 
systematic, semiperturbative approach
to determining the strength and other thermodynamic properties of the
phase transition.  Their idea is the following.  All thermodynamic
properties can be derived in the Matusbara formulism, ie by considering
a Euclidian path integral in which time runs from 0 to $\beta=1/T$ 
with periodic boundary conditions for bosons and antiperiodic boundary
conditions for fermions.  To compute the thermal expectation value
of an operator ${\cal O}$, we find
\begin{eqnarray}
\langle {\cal O} \rangle & = & \frac{\int{\cal D} \Phi {\cal D}
A^{\mu} {\cal O} \exp
(-S)} {\int{\cal D} \Phi {\cal D} A^{\mu} \exp (-S)} \, ,
\\
S & \equiv & \int_{0}^{\beta} dx_0 \int d^3x ( {\cal L} 
+  {\cal L}_{\rm ct} ) \, ,
\nonumber \\
{\cal L} & \equiv & \frac{1}{4g_4^2} F^a_{\mu \nu} F^a_{\mu \nu} +
(D_\mu \Phi)^{\dagger} D_\mu \Phi + m_{H4}^2 \Phi^{\dagger} \Phi +
\lambda_4 (\Phi^{\dagger} \Phi)^2 \, .
\end{eqnarray}
The subscript $4$ means that these values depend on renormalization
point exactly as they do in the 4 dimensional vacuum theory, as do
the wave functions; the theory
also requires counterterms represented by ${\cal L}_{\rm ct}$.
(For simplicity we have not written terms involving fermions, 
but they should be present.)  
By Fourier transforming the time direction, one obtains a
3 dimensional theory in which the Fourier components with nonzero
Matsubara
frequency appear as a Kaluza-Klein tower of massive modes. 
Following \cite{Landsmann},
they construct an infrared effective theory of the zero 
frequency mode,  integrating out the massive 
modes by computing a set of correlators in each theory and 
matching them in the infrared.
Provided that the operator ${\cal
O}$ consists of spatially extended, equal time combinations of bosonic
operators, its expectation value is then well approximated by
\begin{eqnarray}
\langle {\cal O} \rangle & = &  \frac{\int{\cal D} \Phi {\cal D}
A^{\mu} {\cal O} \exp
(-\beta H)} {\int{\cal D} \Phi {\cal D} A^{\mu} \exp (-\beta H)} \, ,
\\
\beta H & \equiv & \beta \int d^3x \frac{1}{4g^2}F_{ij}^a F_{ij}^a +
(D_i \Phi)^{\dagger} D_i \Phi + \frac{1}{2} D_iA_0^a D_iA_0^a +
\frac{g^2}{4} A_0^2 \Phi^{\dagger} \Phi + 
\nonumber \\ & & 
\frac{m_{D0}^2}{2} A_0^2 +
m_{H0}^2 \Phi^{\dagger} \Phi + \lambda (\Phi^{\dagger} \Phi)^2 \, .
\label{dimredH}
\end{eqnarray}
Here the wave functions and couplings do not renormalize but have
their values fixed (at a given temperature) by the matching process.
(We have dropped an extremely small $A_0^4$ term and a slight 
correction to the coefficient in the $A_0^2 \Phi^{\dagger} \Phi$ term,
as in \cite{FKRS}.)  The mass terms, on the
other hand, do depend on the renormalization procedure, as indicated
by the subscript $0$, meaning the bare values.  They are related to
the renormalized values, determined in the perturbative matching
procedure, by
\begin{equation}
m_{D}^2 (\mu) = m_{D0}^2 + \delta m_D^2 ( \mu ) \, ,
\end{equation}
and similarly for $m_H^2$.  In regulations where linear divergences do
not vanish, such as lattice regulations, the counterterm is
substantial and positive, so the bare mass squared may need to be small or
negative.

We have deliberately used different notation than Farakos et. al. to
emphasize that the path integral is over an action which looks like
$H/T$, with $H$ ``almost'' the Hamiltonian of the classical theory.

To investigate the relationship between the dimensionally reduced 
theory above and the thermodynamics of the classical bosonic theory,
we follow a line of reasoning developed in \cite{AmbKras}.
In the classical theory, thermodynamics are described by the partition
function 
\begin{eqnarray}
Z & = & \int {\cal D} A_i^a {\cal D} E_i^a {\cal D} \Pi {\cal D} \Phi
\delta \left( (D_i E_i)^a + g^2 {\rm Re} \Pi \frac{i\tau^a}{2} \Phi \right)
\exp(-\beta H) \\
H & = & \frac{1}{g^2} \left( \frac{1}{4} F_{ij}^a F_{ij}^a +
\frac{1}{2} E_i^a E_i^a \right) + \Pi^{\dagger} \Pi + (D_i
\Phi)^{\dagger} D_i \Phi + 
\\ & &
m_{H0}^2 \Phi^{\dagger} \Phi + \lambda
(\Phi^{\dagger} \Phi)^2 \nonumber \, ,
\end{eqnarray}
where the delta function enforces Gauss' law.  It can be written by
introducing an integration over a Lagrange multiplier $A_0^a$ and
adding to the Hamiltonian 
\begin{equation}
i A_0^a \left( \frac{(D_i E_i)^a}{g} + g{\rm Re} \Pi \frac{i \tau^a}{2} \Phi
\right) \, .
\end{equation}
The measure for $E_i^a$ and $\Pi$ is now trivial and the 
integrals are Gaussian.  Performing them, the partition function becomes
\begin{eqnarray}
Z & = & \int {\cal D} \Phi {\cal D} A_{\mu} \exp(-\beta H) \\
H & = & \int d^3x \frac{1}{4g^2}F_{ij}^a F_{ij}^a +
(D_i \Phi)^{\dagger} D_i \Phi + \frac{1}{2} D_iA_0^a D_iA_0^a +
\frac{g^2}{4} A_0^2 \Phi^{\dagger} \Phi + 
\nonumber \\ & & 
0 A_0^2 +
m_{H0}^2 \Phi^{\dagger} \Phi + \lambda (\Phi^{\dagger} \Phi)^2 \, ,
\end{eqnarray}
identical to Eq. (\ref{dimredH}) except that $m_{D0}^2$ is forced to
be zero.  The actual Debye mass squared then equals the
counterterm.  In lattice regulation, the counterterm turns out to be
\cite{FKRS}
\begin{equation}
\label{latticeDmass}
\delta m_D^2 \simeq \frac{5 g^2 \Sigma T}{4 \pi a} \, , \qquad 
\Sigma = 3.17591 \, ,
\end{equation}
which grows linearly with $1/a$. One choice is
to set the lattice spacing $a$ so that this is actually the
correct Debye mass, but this is not essential. All
that is really needed is that the screening be efficient enough 
that the influence of the $A_0$ field on the infrared physics should be
perturbatively computable (which is the case for the physical value
of $m_D$, and is true for the value in Eq. (\ref{latticeDmass}) for
reasonable values of $a$).  In this case we can use the results of
\cite{FKRS} to relate the thermodynamics of the lattice system to the
thermodynamics without the $A_0$ field, which in turn can be 
related to the thermodynamics with the $A_0$ field and the appropriate
Debye mass.  Except for the need for this correction, an
investigation of the thermodynamics of the classical theory is
equivalent to an investigation of the dimensionally reduced theory.

The preceding discussion 
suggests a profound connection between the thermodynamics of
the classical theory and
the nonperturbative infrared physics of the quantum theory.  To
explore whether the connection extends to real time properties, we
will briefly investigate the real time perturbative
expansion of each theory.

Let us start with classical field theory. For simplicity we 
shall take as usual the example of $\lambda \phi^4$
scalar field theory, and we assume some
regularization, like lattice regularization, is present.
The classical thermal ensemble is 
defined as:  
\begin{eqnarray}
\langle {\cal O} \rangle_\beta & = & \int {\cal D} \phi_i {\cal D} \pi_i 
\exp(-\beta H(\pi_i,\phi_i)) {\cal O} \over { \int {\cal D} \phi {\cal D} \pi 
\exp(-\beta H(\pi_i, \phi_i))}
\label{classens}
\end{eqnarray}
where ${\cal O}$ is some quantity of interest. The
classical field
$\phi_i({\bf x})$ and its momentum $\pi_i({\bf x})$ are those at
some particular initial time $t_i$: the measure is to be thought
of as a measure on the space of initial conditions for the field.
The Hamiltonian $H$ is $H_0+H_{int}$, with 
$H_0=\int d^3 x {1\over 2} (\pi_i^2+ 
\phi_i^2)$ and $H_{int}= {1\over 24} \int d^3x \lambda \phi_i^4$.
Since both the measure and 
the 
Hamiltonian are time independent, the expectation value $
\langle {\cal O} \rangle_\beta$ is independent of $t_i$. As far as 
${\cal O}$ is concerned, it may be any function of the classical
field and momentum, evaluated at any time. For example ${\cal O} 
= \phi({\bf x}, t) \phi({\bf 0}, 0)$ gives the classical unequal time
two point correlator, where $\phi({\bf x}, t)$ is the 
classical solution defined by the initial conditions 
$\phi_i({\bf x})$ and 
$\pi_i({\bf x})$ at $t=t_i$.

Solving the classical theory perturbatively is straightforward.
If one is interested in the correlator 
$\langle \phi({\bf x}, t) \phi({\bf 0}, 0)
\rangle$, for example, one can choose $t_i=0$ and 
solve the classical field equation 
\begin{eqnarray}
(\partial_t^2 -\nabla^2+ m^2) \phi = -{1\over 6} \lambda \phi^3
\label{classeq}
\end{eqnarray}
with an integral equation:
\begin{eqnarray}
\phi(x, t) = \phi_{free}(x, t) -\frac{\lambda}{6} \int_0^t dt' 
\int d^3x' G_R(x-x',t-t') \phi^3(x', t') \, .
\label{classeqsol}
\end{eqnarray}
where $ \phi_{free}$ is a solution of the free theory,
\begin{equation}
G_R(x-x',t-t') = \int \frac{d^3 k}{(2\pi)^3}  e^{i{\bf k \cdot (x-x')}} 
\frac{ \sin (\omega_k (t-t'))}{\omega_k}
\end{equation}
is the retarded Greens function, and $\omega_k = \sqrt{k^2+m^2}$. 
The iteration of this equation produces the solution for 
$\phi(x, t)$ to all orders in $\lambda$. 
Of course 
$\phi_{free}(x, t)$ is easily expressed in terms of 
$\phi({\bf x}, 0)$ and 
$\pi({\bf x},0)$:
\begin{equation}
\phi_{free}(x, t) = 
\int {d^3k\over (2\pi)^3} e^{i{\bf k.x}} \left( \phi({\bf k},0){\rm cos}
(\omega_k t) + \pi({\bf k},0){\rm sin}
(\omega_k t)/\omega_k \right) \, .
\label{classosol}
\end{equation}
To evaluate (\ref{classens}) one 
expands $e^{-\beta H_{int}}$ in powers
of $\lambda$, and performs the Gaussian integrations over 
$\phi({\bf x},0)$ and
$\pi({\bf x},0)$ occurring at each order in $\lambda$.
These integrals are 
summarized by the generating functional:
\begin{eqnarray}
\langle e^{\int d^4 x \phi_{free}(x) J(x)} \rangle_{\beta,free} &=&
e^{{1\over 2} \int d^4 x \int d^4 x' J(x) G(x,x') J(x)} \\
G(x,x')\equiv <\phi_{free}(x) \phi_{free}(x')> 
&=&T\int {d^3 k\over (2\pi)^3}  e^{i{\bf k \cdot (x-x')}} 
\frac{ \cos (\omega_kt )}{\omega_k^2} \, .
\label{classgenfun}
\end{eqnarray}

To summarize, the interactions occur in
two places: first, in 
$e^{-\beta H}$, which defines the thermal state and determines the 
equal time correlators, and second, in the classical evolution
of the fields between the times of interest in unequal time
correlators. The same is true in the quantum theory.

In the quantum theory, one wants to compute the quantum expectation
values of operators: 
\begin{eqnarray}
\langle {\cal O} \rangle_\beta & = & {{\rm Tr}( e^{-\beta H} {\cal O} )
\over {\rm Tr}( e^{-\beta H})}
\label{qmens}
\end{eqnarray}
where the trace is over any complete set of states, 
and ${\cal O}$ is the product of field operators, expressed in 
terms of 
Heisenberg fields. As in the classical theory, one can
evaluate this expression in an interaction picture 
defined at any time (e.g. $t=0$).
Rewriting (\ref{qmens}) in terms 
of the interaction picture operators $\phi^I({\bf x}, t)$ and
$U(t) = Te^{-{i \over \hbar} \int_0^t dt' H^I_{int}(t')}$, we have 
\begin{eqnarray}
{\rm Tr}\left( e^{-\beta H} \frac{1}{2}(\phi^H({\bf x}, t) 
	\phi^H({\bf 0}, 0) + \phi^H({\bf 0}, 0) \phi^H({\bf x}, t)) \right)
	\qquad \qquad \qquad \qquad \nonumber \\ = 
	\frac{1}{2}\sum_i e^{-\beta E_i} \langle i|U(-i\hbar \beta) 
	U^\dagger(t) \phi^I({\bf x}, t)
	U(t) \phi^I({\bf 0},0) |i\rangle + {\rm c. \; c.}
\label{qexp}
\end{eqnarray}
where the sum is over a complete set of interaction picture occupation 
number states. This expression is straightforward to evaluate,
by expanding all the $U$ operators 
in powers of $\lambda$ and then 
evaluating the ensuing free field correlators. 

To see the connection with the classical theory, 
note that $ U^\dagger(t) \phi^I({\bf x}, t)
U(t)$ is just the original Heisenberg field, which obeys the
classical field equation (\ref{classeq}), and as in the classical 
theory, this 
may be solved by iterating the integral equation (\ref{classeqsol}),
with the identical retarded Greens function. The perturbative 
expansion of the Heisenberg operator yields this Greens function via
$[\phi^I(x, t), \phi^I(x,t') ] = i \hbar G_R(x-x',t-t'), t>t'$;
the $\hbar $ cancels the $\hbar^{-1}$ in the time evolution
operator, to all orders in $\lambda$. 
So the `dynamical' part of the calculation of unequal time correlators 
in the classical 
and quantum theories are actually identical\footnote{except that one must be
careful about operator ordering in using the Heisenberg operator
equation of motion.  The products of operators one gets will not be
completely ordering averaged.  However, re-arranging their order only
introduces commutators which are $O(\hbar)$, which by dimension counting 
means $O(\hbar \omega/T)$.  It turns out that when one asks 
questions about ordering averaged products of operators, as we do here,
these re-ordering corrections actually first appear at second order
in $\hbar$, see \cite{Bodekernew}.}. It is also
remarkable that when organized this way,
both theories have exactly the
same set of Feynman diagrams, it is just the Feynman rules that
are different.

Differences arise in 
two places. First, in 
the expectation values of the resulting series of free fields.
In the quantum theory the generating function is given as in 
(16) but with $G_{class}(x,x')$ replaced by 
\begin{eqnarray}
G_{quantum}(x,x')&=&\int {d^3 k \over (2\pi)^3}
e^{i{\bf k.(x-x')}} {\hbar \over \omega_k} 
{\rm cos} (\omega_kt ) \left[ \frac{1}{e^{\beta \hbar \omega_k }
-1} + \frac{1}{2} 
\right] \, .
\label{qgenfun}
\end{eqnarray}
(This `thermal Wicks theorem' has been discussed recently by
\cite{evans} - the simplest way to derive it is to note that 
the relevant path integral is Gaussian, from which (16) 
follows. The `thermal Wicks theorem' is true for any
initial density matrix which is Gaussian.).
The second difference arises because 
$U(-i \hbar \beta) = T e^{ -{1\over \hbar} \int_0^{\hbar \beta} H_{int}
(\tau)}$ 
is not exactly equal to
$e^-{\beta H_{int}(0)}$. This difference can be attributed to
the difference between the quantum and classical (or `dimensionally
reduced') thermal states.

We now see when the classical and quantum 
results agree. For low frequency bosonic 
modes at high temperatures
$\hbar \omega /T \ll 1$, the occupation number is high and 
the bracketed expression in Eq. (\ref{qgenfun}) can be
expanded in an asymptotic series,
\begin{equation}
 \left[ \frac{1}{e^{\beta \hbar \omega_k }
-1} + \frac{1}{2} \right] = \frac{T}{\hbar \omega_k} + \frac{1}{12} 
	\frac{\hbar \omega_k}{T} + \ldots
\end{equation}
and the leading term precisely reproduces
the classical Greens function. 
Note that the usual $\hbar$ in the propagator is
cancelled by a $\hbar^{-1}$ in the occupation number. 
Second, as noted above, $H_{int}^I(\tau)$ is not
equal to $H_{int}^I(0)$ - the imaginary time dependence
of the free field operators causes a difference in the
thermal state corrections.
However $H_{int}^I(\tau)$ can be expanded as a Taylor
series in $\tau$, and the $\tau$ dependent corrections  come
in the form $\hbar \omega_k \tau < \hbar \omega_k/T$.  Thus again 
for $\hbar \omega /T \ll 1$, the corrections are small
\footnote{In fact for equal time correlators of $\phi$ only, 
these thermal state corrections occur only at order 
$(\hbar \omega_k/T)^2$, as is seen by writing 
$U(-i\hbar \beta) \simeq e^{+\beta H_0} e^{-\beta H}$ with
(quantum) corrections arising from
commutators of $H_0$ with $H_{int}$. At first order
in $\hbar$ there is only the single commutator, which is odd in
$\pi$ and therefore does not contribute to the correlator.}.

To summarize, as long as one is interested in phenomena
which are dominated by low frequency bosonic 
modes, for which $\hbar \omega /T \ll 1$, the 
the classical field theory provides a good description. 
Note that the thermal classical theory contains much more than just 
tree diagrams - it sums up all loop diagrams as well, 
in the approximations  first that the quantum propagator is replaced
by the classical ($\hbar$ independent) piece and second that
 the thermal state is taken to be that of the 
`dimensionally reduced' three dimensional theory.

The above argument works perfectly in a regularized (e.g. lattice)
field theory.
The only catch in applying it to a continuum
field
theory is that the condition $\hbar \omega /T \ll 1$ cannot be true for
the very high $k$ modes.
For large $k$ the quantum and classical propagators
are very different, and
the ultraviolet behavior of the theory approaches that of 
the vacuum quantum
theory.  However, if we are interested in long wavelength correlation
functions, these ultraviolet modes will only emerge in diagrams
in which there are a few closely spaced vertex insertions, which we
should be able to replace with an expansion in local operators.
The classical theory will only make
sense as a regulated, infrared effective theory with a Lagrangian
which takes into account these local, quantum effects.  For instance,
the couplings will be renormalized, to a scale given roughly by the
temperature.

In addition to short distance vacuum corrections, which simply lead
to the usual renormalization of the couplings to a scale given
roughly by $T$, there are thermal (nonvacuum) short range corrections,
whose form is not restricted by Lorentz invariance because the thermal
bath chooses a preferred time direction.  There has been intense
research into how these effects influence the behavior of the infrared
(soft) modes, and it has been shown that the parametrically leading
effects can be summarized as a set of ``hard thermal loop'' effects
which can be incorporated into the Lagrangian of the infrared theory
\cite{BraatenI}.  

For this reason, Bodeker et. al. proposed to include the hard thermal loop
effects in numerical investigations of the classical field theory
\cite{Smilga}.  In fact, their work shows that they are already included,
because the most ultraviolet classical excitations in a (lattice cut off)
classical simulation perform the same role as the high frequency
thermal excitations in the quantum theory; all that is
different from the quantum theory is the shape of the cutoff and the
total strength of the hard thermal loop effects.  The total strength 
of the hard thermal loop effects generally cancels in calculations of
dynamical properties of the plasma \cite{BraatenII}, because the hard
thermal loops represent both a bath of particles against which to scatter,
and a bath of particles which screen interactions so that only shorter
range scatterings can take place.  It is not generally believed that the
total number of particles available to contribute to the hard thermal
loops affects the infrared dynamics, and if it does, then dynamical
properties would show a power law in $a$ (the lattice spacing) dependence.
Testing for a small $a$ limit to dynamical properties therefore constitutes
a check that the magnitude of hard thermal loops does not matter, or
at least that the dynamical property in question converges to a limit
in the (parametrically justified) limit of
large hard thermal loop contributions.  The only remaining concern is
that the functional form of the hard thermal loops is incorrect because
of cutoff artefacts; but by varying the form of the cutoff one can test
for this as well, and at least in the case of the motion of Chern-Simons
number it appears that lattice artefacts are small and consistent
with zero \cite{Moore2}.

Hence, we conclude that the classical theory can be used to examine
the thermodynamics of Yang-Mills Higgs theory and
allows the study of the infrared dynamics as well, though here there
are some legitimate concerns involving hard thermal loops.

\section{Numerical implementation of the theory}
\label{Numerics}
	
The numerical implementation of classical, real time Yang-Mills Higgs
theory is well developed in the literature
\cite{Kogut,Ambjornetal,AmbKras,Moore2}.  The implementation we
use here is identical to that of \cite{Ambjornetal}; we consider
the theory at zero Weinberg angle (no $U(1)_Y$) and take as
degrees of freedom an SU(2) matrix on every link of a 3+1 dimensional
lattice which is toroidal in the 3 space directions but infinite
in the time direction, and a fundamental complex scalar at each 
vertex.  The lattice action is
\begin{eqnarray}
\beta_L S & = & \beta_L \bigg[ 
- \sum_{\Box_s} 1- \frac{1}{2} {\rm Tr} U_{\Box_s}
   +\frac{1}{(\Delta t)^2}\sum_{\Box_t} 1-\frac{1}{2} {\rm
   Tr}U_{\Box_t} 
\nonumber \\
& - & \sum_{x,t} \sum_i \frac{1}{2}(\Phi(x,t)-U_i(x)\Phi(x+i,t))^{\dagger}
   (\Phi(x,t)-U_i(x)\Phi(x+i,t)) 
\nonumber \\
& + & \frac{1}{(\Delta t)^2} \sum_{x,t}
   \frac{1}{2}(\Phi(x,t)-U_0(x)\Phi(x,t+\Delta t))^{\dagger}
   (\Phi(x,t)-U_0(x)\Phi(x,t+\Delta t)) 
\nonumber \\
& - & \sum_{x,t} (\frac{m_{H0}^2}{2}
   \Phi^{\dagger} \Phi + \frac{\lambda_L}{4} (\Phi^{\dagger} \Phi)^2) 
	\bigg] \, .
\end{eqnarray}
We absorb the gauge coupling into the lattice temperature $\beta_L$,
which is related to the lattice spacing and the continuum
temperature through
\begin{equation}
\beta_L = \frac{4}{g^2 a T} \, .
\end{equation}
We also give the Higgs fields the same wave function normalization as
the gauge fields, which is natural and computationally convenient.
The Higgs field is treated as four independent real entries, and the
relation between the lattice value and the continuum one is
\begin{equation}
\beta_L^2 \Phi^{\dagger} \Phi = \frac{4}{g^2} \phi^2_{\rm cont} =
\frac{8}{g^2} \Phi^{\dagger} \Phi_{\rm cont}  
\end{equation}
for the normal continuum definitions of $\phi$ and $\Phi$.
With this wave function normalization, the 
scalar self-coupling $\lambda_L$ is related to the usual
continuum one by 
\begin{equation}
\lambda_L = 4 \lambda/g^2 \, ,
\end{equation}
which is parametrically order 1 if $\lambda \sim g^2$, as is natural
from the renormalization structure of the theory.  The only small
numbers in the classical theory are $1/\beta_L \propto a$ the lattice
spacing, and $m_H^2$ the renormalized Higgs mass squared.

Varying the action with
respect to the link in the time direction generates a constraint,
analogous to the continuum condition $D_i E_i^a/g^2 = 
{\rm Re}(\Pi^{\dagger} i \tau^a \Phi) = \rho$, which is Gauss's law; 
variation with respect to the space links and the Higgs field
components give equations of motion which allow all future times to be
determined from two neighboring time slices (provided that these
initial conditions satisfy the constraints), once an ambiguity in the
time evolution, due to the freedom to change gauge independently at
each spacetime
point, has been used up by choosing the gauge which always makes the
time links $U_0 = I$ the identity.

In practice we keep track of the values of fields on one time slice,
and ``momenta'' $\Pi(t-\Delta t/2) = \Phi(t) - \Phi(t-\Delta t)$ and
$E_i^a(t-\Delta t/2) = -1/2 {\rm Tr} i \tau^a U_i(t) U^{\dagger}_i(t -
\Delta t)$ (after setting the $U_0$ to $I$); as long as $E$ is small, 
the latter relation can easily be inverted to update $U$.

We thermalize the system with the algorithm developed in
\cite{Moore1}.  That is, beginning from an arbitrary initial
condition, we repeatedly draw the momenta from the correct thermal
distribution (achieved by choosing them as Gaussian random variables,
and then orthogonally projecting to the constraint surface) 
and then evolve the system under the equations
of motion for some time, allowing the thermalization to mix with the
coordinates $U$ and $\Phi$.  The momenta are then discarded and the
procedure is repeated, as many times as desired.  The algorithm has
a time stepsize ambiguity, which we handle as in \cite{Moore2}; the
thermalization is only accurate to $O((\Delta t)^2)$, a level which is
sufficient because the evolution algorithm is also inaccurate at this
level.  In this work we always use $\Delta t = 0.05$ in lattice units,
which is sufficient to hold stepsize systematics below statistical
errors.

As discussed above, a thermalization algorithm for classical
Yang-Mills Higgs theory can be considered a canonical ensemble
(fixed temperature)
Monte-Carlo algorithm for the
dimensionally reduced theory.  (In fact, except for the Gauss
constraint, it exactly resembles the hybrid algorithm of Euclidean 
lattice gauge theory.)  The algorithm is very efficient at
exploring the thermodynamics of one phase; but it is very bad
at going between the two phases.  This is because it must move
smoothly from one phase to the other, and mixed phase configurations 
contain phase boundaries which
have positive surface tensions.  In the limit of large box size, the
suppression of such configurations grows roughly as
$\exp(-2 L_1 L_2 / \sigma)$, with $L_1$ and $L_2$ the two shortest lengths
of the box and $\sigma$ the surface tension.  It is
necessary to change phase several times to get good statistics on the
free energy difference of the two phases, and hence to determine the
critical temperature, but as the box size grows, it will become
essentially impossible for the canonical evolution algorithm to do so;
it will only be capable of thoroughly
exploring one minimum, but not of comparing the two.  For this 
reason, the literature generally considers a multicanonical ensemble (in
which a global reweighting term is added to make mixed phase
configurations more favorable but is then accounted for in computing
thermal averages of operators) a more powerful technique for 
exploring the thermodynamics of first order phase transitions.

\begin{figure}
\centerline{\psfig{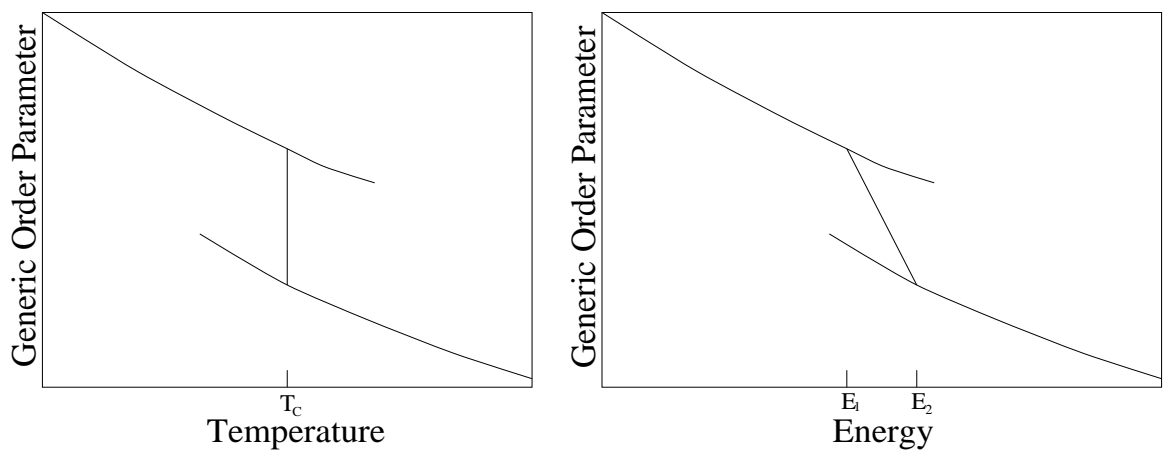}}
\caption{\label{f1} Schematic depiction of the dependence of an
order parameter on temperature and energy, including metastability
lines and the equilibrium coexistence line.  Though equilibrium phase
coexistence only occurs precisely at $T_c$, it happens for a range of
energies between $E_1$ and $E_2$.  Phase mixture is stable for a
fixed energy system if the energy is in this range.}
\end{figure}

In fact, properties of the metastable phases, the equilibrium
temperature, and virtually all other themodynamic properties can be
extracted by using a microcanonical (fixed energy) ensemble.  
A microcanonical ensemble is achieved by thermalizing the system to 
some temperature, but then allowing it to evolve under the equations 
of motion indefinitely, without ever re-randomizing the momenta.  The
system is strongly ergodic, so for general initial conditions it will
thoroughly explore the fixed energy subsurface of phase space.  In the
large volume limit, this would become approximately equivalent to a
canonical ensemble, except that there is a first
order phase transition.  There is a finite range of energies 
where the fixed energy equilibrium configuration is mixed phase, and
if one can find one mixed phase configuration in this range, then the
microcanonical algorithm can use it to thoroughly explore mixed phase
configurations, and in particular to extract the phase transition
temperature and information about the phase interfaces.  Since the
whole range of mixed phase configurations occur at a single
temperature, the canonical ensemble is
not well suited to exploring phase coexistence.  This difference between the
two ensembles is illustrated in Figure \ref{f1}.

To exploit this property of the evolution algorithm, it is necessary
to find a way to measure temperature during a fixed energy
evolution, and to very gradually increase or decrease the energy so
that a range of configurations can be explored.  If there were no
Gauss constraint, it would be easy to measure the temperature.  The
momenta would be true Gaussian degrees of freedom, and their average
energy (averaged over the set of momenta and over time to remove
fluctuations) should obey equipartition.  The Gauss constraint only
mildly complicates this picture.  The constraints are linear in the
momenta, and remove one Gaussian degree of freedom each.  Although
they are not local, so we do not know an orthogonal basis for the
remaining set of independent Gaussian degrees of freedom, we still
know their total number, so the total kinetic energy
should equal (number of degrees of freedom=10)$T/2$.  Averaging
over time removes statistical fluctuations and gives a clean value for
the system temperature.

It is also quite easy to gradually change the amount of energy in the
system.  Our algorithm to do this is to simulate a gradual, adiabatic
expansion or contraction of the lattice.  All coordinates ($U$ and
$\Phi$) are updated as usual, but all momenta $E$, $\Pi$ are multiplied
by the same factor $(1+\epsilon \Delta t)$ each time step.  The system
heats with time constant $1/\epsilon$.  If $\epsilon$ is negative, the
system will cool.  Since the Gauss constraint is linear in $E$ and
$\Pi$, this heating algorithm identically preserves Gauss' Law.  

It is
also sometimes necessary to make the heating or cooling more local, so
that the temperature can be kept uniform even if some phenomenon
liberates heat locally.  To do this we bin the lattice into boxes or
slabs, and measure the temperature in each.  Then each box has its
momenta multiplied using a different $\epsilon$, chosen to drive each
box independently towards some desired temperature.  The algorithm
generates small violations of Gauss' Law on the interfaces between boxes,
which we remove with the orthogonal projection algorithm presented in
\cite{Moore1}.

Note that neither of these heating algorithms will keep the system
in equilibrium; of course no heating algorithm will.  But if they
are applied gradually, the system should remain very close to 
equilibrium (though it will tend to superheat or supercool into
metastable phases, as noted earlier).

\section{Thermodynamic properties}
\label{thermodynamics}

We are now ready to explore the equilibrium properties of Yang-Mills
Higgs theory.  We will not attempt a thorough investigation of the 
strength of the phase transition as a function of $\lambda_L$, as very
accurate calculations already exist \cite{KLRSresults,otherguys}; rather we
will investigate how well the microcanonical technique can be applied
to determining the various properties of the 
phase transition.  Similarly, we have not
attempted to make a small lattice spacing extrapolation; all the data
presented below are for $\beta_L \simeq 8$ or $6$ and $\lambda_L =
0.20$, which, in the notation of \cite{KLRS}, is $x=0.05$.

\subsection{Metastability and hysteresis}

\begin{figure}
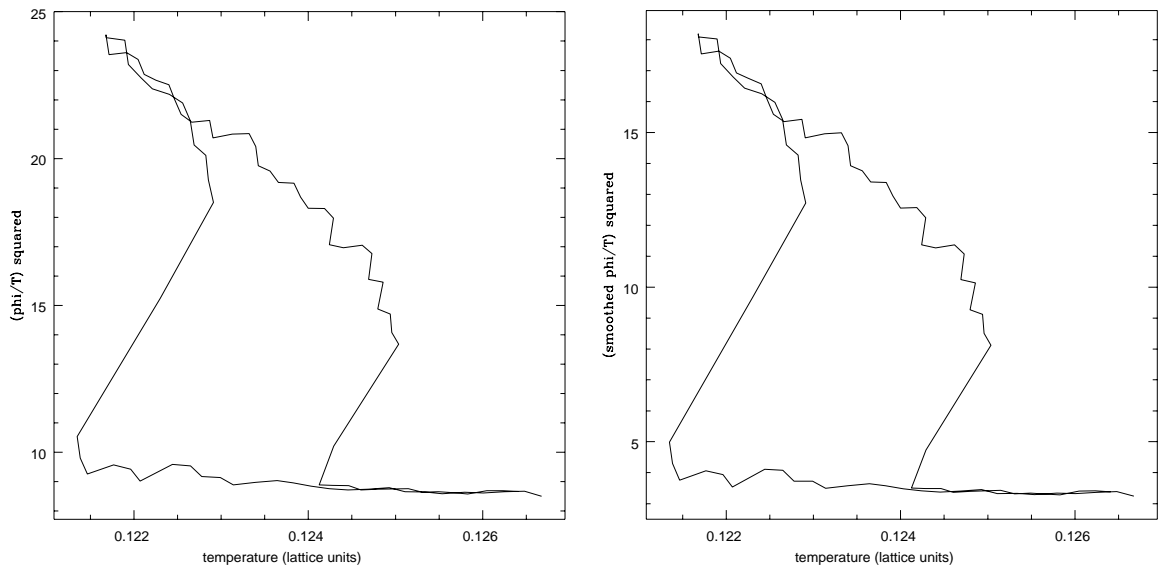
 
\mbox{\psfig{file=phi.epsi,width=2.9in}}
\hspace{0.1in}
\mbox{\psfig{file=phismooth.epsi,width=2.9in}}
\caption{\label{OPjump} Lattice $\Phi^{\dagger} \Phi/T^2$ as a
function of temperature ($1/\beta_L$) for a $30^3$ lattice; at left,
without smoothing, and at right, with one nearest neighbor smoothing.
The system begins in the symmetric phase at
high temperature and is cooled through the phase transition,
and is then heated back to the original temperature.  The
system supercools during the cooling and superheats during the
heating, resulting in a hysteresis loop.  The cooling represents a
total elapsed time of $1280$ lattice units; the heating is twice as long,
with longer bins, to suppress fluctuations.  The fluctuations
in $\Phi^{\dagger} \Phi/T^2$ are much larger in the broken than
the symmetric phase, except near where the symmetric phase becomes
unstable, where the fluctuations become larger.  Both the jump in the
order parameter and the fluctuations are almost identical before and
after smoothing, so both are infrared phenomena; but the symmetric
phase value of $\Phi^{\dagger} \Phi$ is primarily ultraviolet, and
is suppressed by the smoothing.}
\end{figure}

\begin{figure}
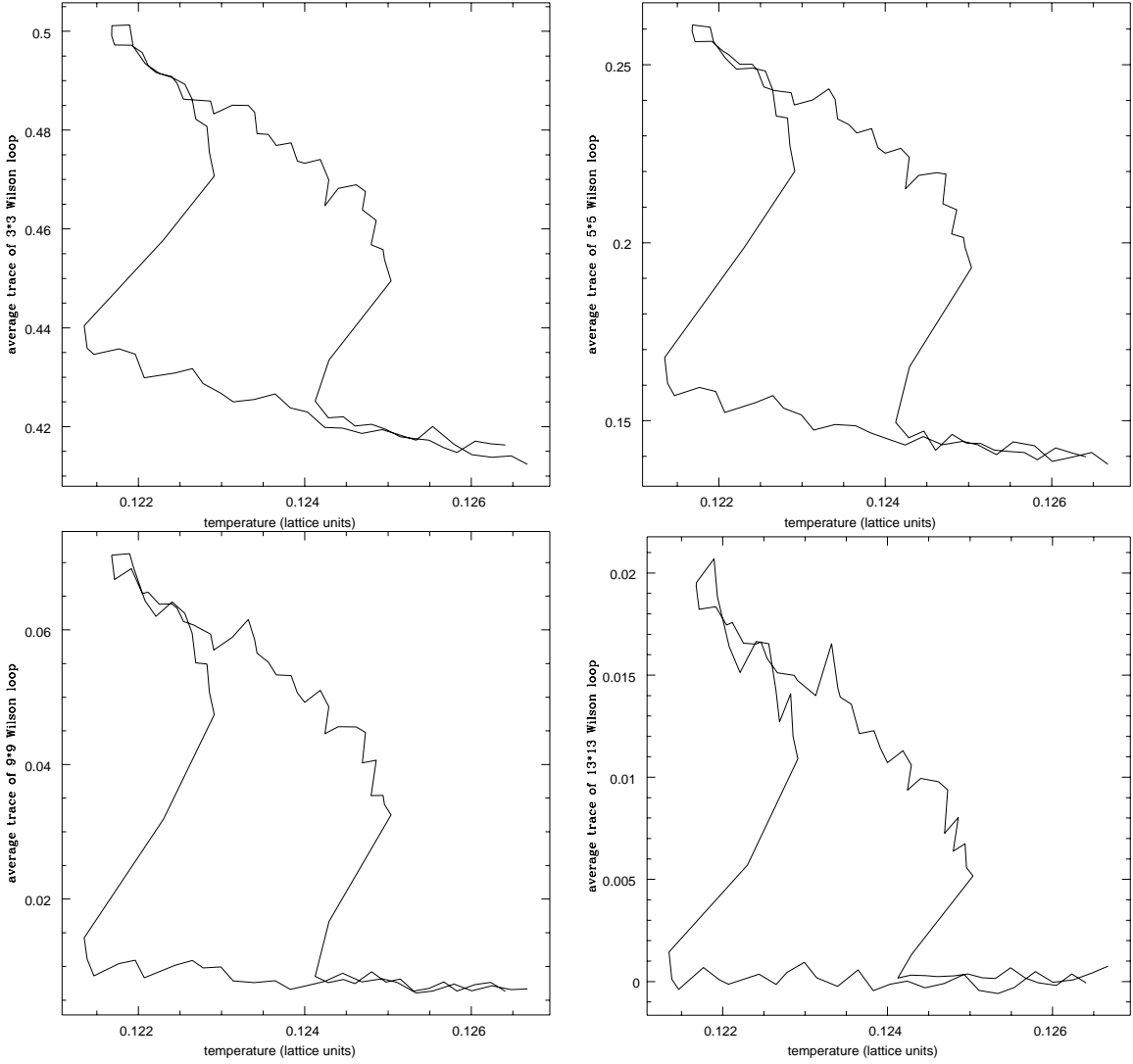

\mbox{\psfig{file=w33.epsi,width=2.9in}}
\hspace{0.1in}
\mbox{\psfig{file=w55.epsi,width=2.9in}}
\mbox{\psfig{file=w99.epsi,width=2.9in}}
\hspace{0.1in}
\mbox{\psfig{file=wdd.epsi,width=2.9in}}
\caption{ \label{Wilsons} Average of the trace of some 
Wilson loops for the same run.  The normalization is such that the
trace is $1$ in vacuum.  Fluctuations in the Wilson loops are
about the same size in each phase, and they follow quite closely the
fluctuations of $\Phi^{\dagger}\Phi$.  For the $9\times 9$ loop the
symmetric phase Wilson loops are almost zero, showing no long range
order; they are zero within tight statatistical error for $13 \times
13$ loops, and here the broken phase values are nearly zero as well.}
\end{figure}

The first thing we can investigate is the temperature dependence of
order parameters, including metastable branches.  We do this by thermalizing
a $30^3$ lattice with bare Higgs mass squared $m_{H0}^2 = -0.3223$ in 
lattice units and $\lambda_L = 0.20$ at a temperature above the 
phase transition temperature, estimated to occur where $m_{H0}^2$ equals the 
one loop counterterm,
\begin{equation}
\delta m_{H}^2 = \frac{(9 + 6 \lambda_L) \Sigma }{ 4 \pi \beta_L } \, , 
\end{equation}
which happens when $\beta_L 
\simeq 8.0$.  The system is gradually cooled, and the lattice 
temperature and order parameters such as $\Phi^{\dagger} \Phi$ are averaged
in time bins longer than the lattice length but much shorter than
the full length of the cooling.  After the order parameter jumps to the
broken phase value, the system is heated back to the original temperature.
As a function of temperature, order parameters exhibit hysteresis, 
returning to the symmetric phase at a higher temperature than they left
it.  This is the signature of a first order phase transition.  The 
hysteresis curve for $\Phi^{\dagger} \Phi$, a once 
smoothed $\Phi^{\dagger} \Phi$ defined in Appendix B,
and the traces of a few sizes of Wilson loops are presented in
Figures \ref{OPjump} and \ref{Wilsons}.  
Smoothing $\Phi$ greatly reduces the contributions from the 
most ultraviolet fluctuations,
but barely touches the infrared fluctuations.  We expect that almost all of
the value of $\Phi^{\dagger} \Phi$ in the symmetric phase should arise
from ultraviolet fluctuations, which should contribute 
$\beta_L \Sigma/\pi \simeq 8.1$ to
$\Phi^{\dagger} \Phi \beta_L^2 $ \cite{FKRS}, so the value of $\Phi^{\dagger}
\Phi$ should be much lower in the symmetric phase after smoothing; but
almost all of the difference between phases should be infrared and
unaffected by smoothing.  Figure \ref{OPjump} verifies this,
and also shows that almost all of the random thermal fluctuations in the
order parameter are infrared, because the details of the fluctuations
in $\Phi^{\dagger} \Phi$ are almost unchanged by the 
smoothing.  Note also that, except near the spinodal point, the fluctuations
in the symmetric phase value of $\Phi^{\dagger} \Phi$ are much smaller than
in the broken phase; this is because of interference between fluctuations
and the condensate in the broken phase.
Also note that there is a large correlation between the fluctuations
of different sized Wilson loops, and between the fluctuations in Wilson
loops and the fluctuations in $\Phi^{\dagger} \Phi$.  

Examining the 
Wilson loop plots, we see that, in the symmetric phase, 
traversing a $9 \times 9$ loop yields an 
almost completely random SU(2) phase, so the symmetric phase is disordered
on the scale of $\beta_L$.  In the broken phase traces of Wilson loops
fall off more slowly, but $13 \times 13$ Wilson loops show almost no order
and it cannot be meaningful to speak of anything in a nontrivial 
representation of SU(2) as having
any correlations beyond this scale.  Hence the box size was abundantly 
larger than the longest possible correlation length and the results should
represent the infinite volume limit.  
We also verified this by repeating the run
on a $20^3$ lattice; the results were the
same within error, but the fluctuations were larger because they were
not averaged over as much four-volume, so we will not present the
results here.

\subsection{Equilibrium temperature}

The results presented in Figure \ref{OPjump} can be used to
determine the jump in $\Phi^{\dagger} \Phi$ at any 
temperature for which both phases are reasonably 
metastable.  However, it cannot be used to determine the nucleation 
rate, because in the early universe the supercooling occurred much more
gradually than in any concievable simulation and the true nucleation
point occurs when the tunneling probability is too small to observe in
a simulation.  It also cannot give us the equilibrium temperature, although
if we knew the equilibrium temperature we could use the hysteresis plot
to find the jump in the order parameter at equilibrium.  To get the 
equilibrium temperature, we must establish phase coexistence in a 
microcanonical evolution and measure the temperature.

\begin{figure}
\centerline{\psfig{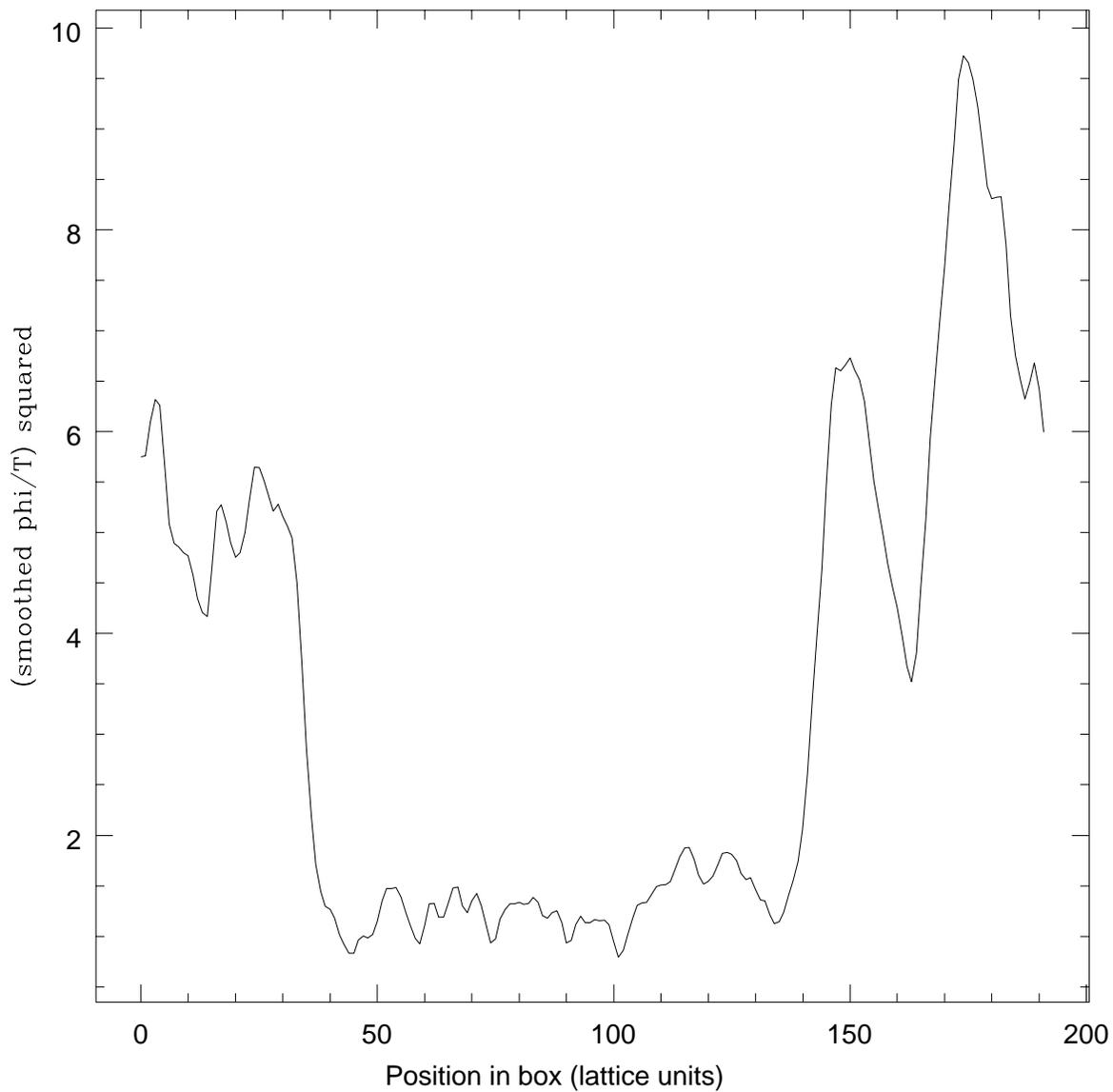}}
\caption{\label{avgij} Six times smoothed Higgs field, averaged over the 
short directions of a $16\times 16 \times 192$ lattice at the equilibrium
temperature $T \simeq 0.1241$, plotted as a function of position in
the box.  The two phases and the phase boundaries 
are clear, although the broken
phase has quite sizeable fluctuations in the order parameter.}
\end{figure}

We do this as follows.  We thermalize an $N \times N
\times 192$ rectangular box with symmetric boundary conditions 
at the same value of $\lambda_L$ and $m_{H0}^2$ used above and $\beta_L
= 8.0$.  Then we apply a small perturbation to $m_{H0}^2$, with an 
amplitude which varies sinusoidally along the long direction of the 
box.  The symmetric phase is favored in one region and the broken phase
is favored in another, so a mixed phase configuration is established.
The perturbation is then slowly removed, and as it is removed the 
system is heated or cooled by a thermostat which tries to balance 
$\int \Phi^{\dagger} \Phi d^3x$ at a value 
intermediate between the two phases.  
Once the perturbation is completely lifted, the system is allowed to
evolve for a long time (at least 500 lattice lengths) without any heating
or cooling to equilibrate fully in the mixed phase configuration and
erase all record of the process by which a mixed configuration was 
generated.  It is then evolved for a long period of time, again without
any heating or cooling, during which the energy in kinetic degrees of 
freedom is averaged and used to establish the 
equilibrium temperature.  We have checked that the box we used was
abundantly longer than that required to contain two domain walls and
a region of each phase, and that the system remained in the two phase
configuration during the whole run.  The two phase nature can be seen
clearly by averaging $\Phi^{\dagger} \Phi$, optionally applying several
iterations of smoothing, over the two short directions
of the lattice and plotting against the long direction, as shown in
Figure \ref{avgij}.  The reason that fluctuations in
phase boundary positions do not drive the system to one or the other
phase is that a fluctuation which expands the broken phase liberates
latent heat, which raises the temperature and makes the symmetric
phase more favorable, and similarly a fluctuation which expands the
symmetric phase absorbs latent heat and makes the broken phase more
favorable.  

We measured the equilibrium temperature in a $16 \times 16 \times 192$
box and in a $32 \times 32 \times 192$ box; the answers are within 
errorbars and average to $\beta_L = 8.059 \pm .002$\footnote{In the
thermalization algorithm, $E(t)$ is drawn from the Gaussian distribution
and $E(t + \Delta t/2)$, the value required for the leapfrog 
algorithm, is determined by a half leapfrog step; but the temperature
quoted is the sum of $\sqrt{1-(\Delta t)^2E^2(t+\Delta t/2)}/\Delta t$,
so the value we quote will have a weak stepsize dependence.}.  
The jump in the 
order parameter $\Phi^{\dagger} \Phi \beta_L^2$ at this 
temperature, read off the hysteresis plot, 
is $9.5 \pm .2$.  This corresponds to a 
jump in the continuum order parameter $\phi$ of $1.54 gT$ ($g$ the weak
coupling).  Two loop perturbation theory predicts a jump of $1.24 g$,
which is smaller\footnote{Here and throughout we use Eq. (34) of 
\cite{FKRS}, with all terms involving $m_L$ dropped, for the two loop 
effective potential}; this might at first seem surprising, considering 
that the values found by Kajantie et. al. 
\cite{KLRSresults} for $\lambda_L$ on either side of
$\lambda_L = 0.2$ are closer to the two loop perturbative value.  The
reason is that they perform an extrapolation to zero $a$ (infinite
$\beta_L$) based on data at several values of $\beta_L$, whereas the 
result we quote above is for one value of $\beta_L$ and contains finite
lattice spacing artefacts.  Because $\lambda_L$ is small and the jump in
the order parameter is quite sensitive to its value, the most important
of these effects may be those which shift the effective value
of $\lambda_L$.  One arises because
we have $A_0$ fields with finite Debye mass squared; if we integrated
them out, then they would shift $\lambda_L$ by \cite{KLRS}
\begin{equation}
\delta \lambda = \frac{-3g^4 T}{128 \pi m_D} \quad \Rightarrow
\quad \delta \lambda_L = \frac{-3}{8 \sqrt{5 \pi \beta_L \Sigma}}
\simeq -0.0188 \sqrt{8/\beta_L}  \, .
\end{equation}

There is also a linear in $a$ correction to $\lambda_L$ arising from
one loop diagrams.
The one loop contribution to the effective potential is 
\begin{equation}
V_1(\phi_0) = \sum \int_{m(\phi=0)}^{m(\phi = \phi_0)} m I(m) dm \, , 
\end{equation}
where the sum is over massive degrees of freedom and $I(m)$, the 
one loop $3d$ lattice tadpole graph, is computed in \cite{FKRS},
\begin{equation}
I = \frac{\Sigma}{4 \pi a} - \frac{m}{4 \pi} - \frac{\xi a m^2}{4 \pi} 
+O(a^2) \, , \quad \xi = 0.1529\, .
\end{equation}
The leading, $1/a$ term generates the linearly divergent mass squared
correction, the next term gives the negative cubic term which determines
the order of the phase transition, and the $a m^2$ term produces an $O(a)$
correction to $\lambda$ which, summed over degrees of freedom, is
\begin{equation}
\delta \lambda = \frac{- \xi aT}{4 \pi} \left( \frac{9 g^4}{16} 
+ 12 \lambda^2  \right) \quad \Rightarrow \quad
\delta \lambda_L = \frac{- \xi}{4 \pi \beta_L} ( 9 + 12 \beta_L^2 ) \, ,
\end{equation} 
which equals $-0.014$ for $\beta_L = 8$.  Hence, the simulations 
described correspond to $\lambda_L$ of roughly $0.167$ rather than $0.2$,
and the two loop perturbative estimate for $\phi$ is $1.40 g$, quite
close to the actual value.

Note that accounting for finite $a$ shifts in $\lambda_L$ as above does
not completely remove finite $a$ or even all linear in $a$ errors, because 
we have not corrected wave functions or removed high dimension operators.
It is probably not profitable to pursue high statistics calculations 
until the required corrections have been computed.

Also note that the jump in $\Phi^{\dagger} \Phi$, together with the
value for $m_{H0}^2$, determines the latent heat, as shown in 
\cite{FKRS}.  A more direct measure of the latent heat is the temperature
change during the spinodal jump from one phase to the other, which
can be looked up on the hysteresis plot (though it must be remembered that
the heating or cooling of the system continued during the jump).
The value obtained from the jump in temperature is consistent with 
the latent heat obtained from the jump in $\Phi^{\dagger} \Phi$.

\subsection{Surface tension}

Another interesting property of the equilibrium system is
the surface tension of the phase interface.  
The surface tension can be determined from the power spectrum 
of fluctuations of the interfaces.  To understand this, consider a surface
with a very large surface tension; since it is taut, it should be flat.  
But for finite
surface tension, the entropy associated with having nonzero fluctuations
on the surface prevents it from being perfectly flat.  For long wavelength
fluctuations, the fluctuation amplitude is small compared to the wavelength,
and different fluctuations approximately decouple; on average they are 
populated according to equipartition.  Thus the infrared limit of the
power spectrum (square of the Fourier coefficients) of the bubble surface
determines the surface tension.  The details are given in Appendix B, where 
we also discuss how we define the bubble wall surface.

To apply the technique discussed in the appendix, 
we evolved an equilibrated mixed phase configuration
for on order 1000 lattice lengths, recording both bubble wall surfaces 
every 2 lattice lengths of time.  Each surface is Fourier transformed,
and the square of each Fourier coefficient is averaged over the run.
For each value of $n^2$ there are several independent coefficients
(from the real and imaginary parts of one or more 
Fourier coefficients from two walls),
and we average these and take the error bars to be the standard 
deviation of the mean.

\begin{figure}
\centerline{\psfig{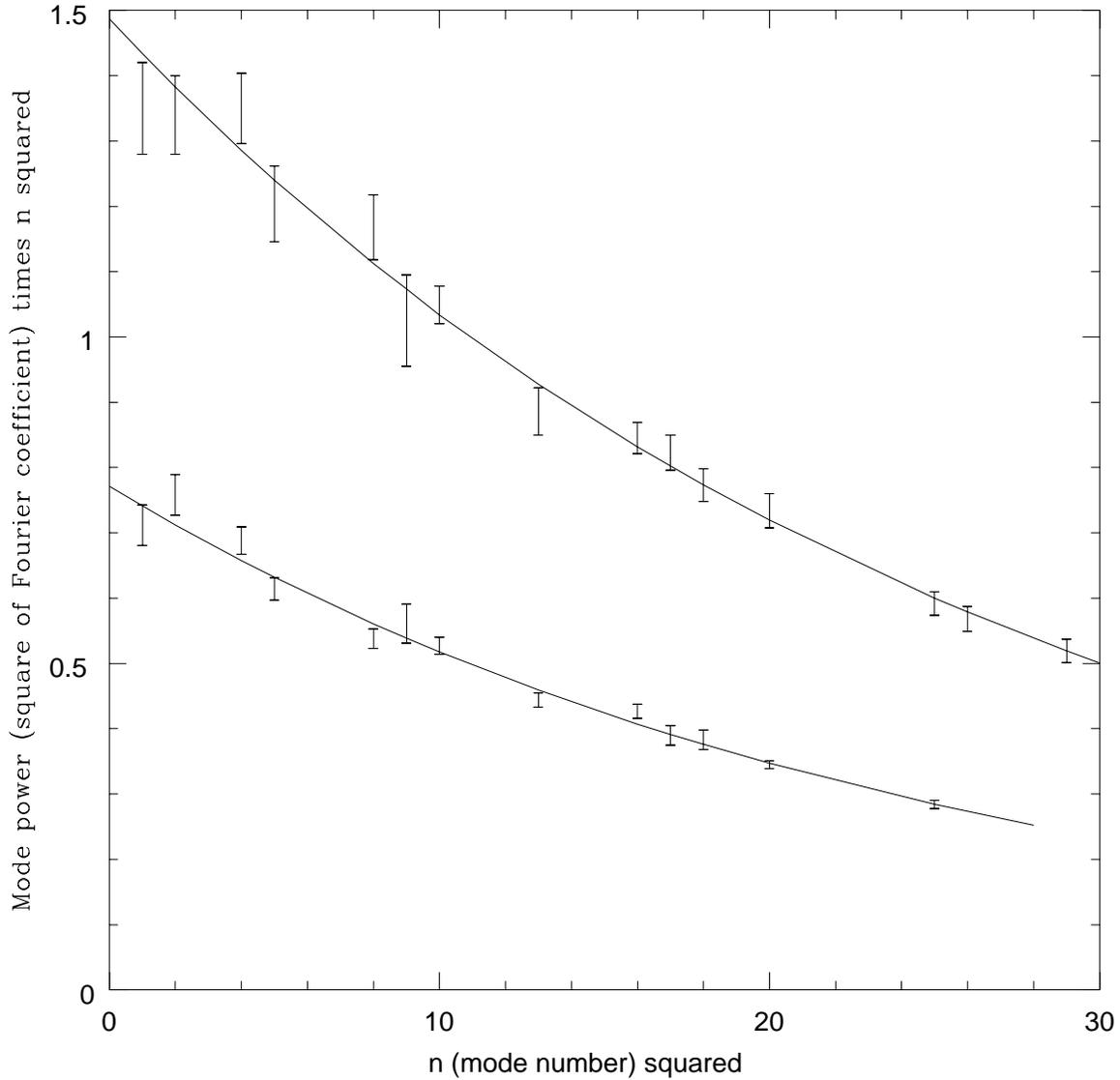}}
\caption{\label{WallPower} Bubble wall surface power spectrum:  
$\langle \tilde{z}_n^2 n^2 \rangle$ plotted against $n^2$.  The upper 
data are for a $32 \times 32$ cross section box at $\beta_L = 8$ and 
the lower data are for a $36 \times 36$ cross section box at 
$\beta_L = 6$.  The fit functions are exponentials, used to extrapolate
the data to the infrared $(n=0)$ limit.}
\end{figure}

We plot the resulting power
spectrum, multiplied by $n^2$ the square of the Fourier mode number,
for a $32 \times 32 \times 192$
lattice at $\beta_L \simeq 8$ and for a $36 \times 36 \times 144$
lattice at $\beta_L \simeq 6$ in Figure \ref{WallPower}.  This is not
a log-log plot; the departure from $1/n^2$ behavior in going from $n^2=1$
to $n^2=25$ is only on order a factor of two.  Part of this departure from
strict power law behavior may arise from the smoothing involved in
defining the bubble wall surface, and some of it may represent interactions
between the high frequency modes on the wall.  To make the infinite
wavelength extrapolation we fit the data with an exponential; the
fit has $\chi^2/\nu$ of $0.67$ and $1.4$ for the $32^2$
and $36^2$ cross section cases and yields surface tensions in physical
units of $.0681 \pm .0009 g^4 T^3$ and 
$.0739 \pm .0008 g^4 T^3$ respectively.  
Note that for $\beta_L=6$ the
finite $a$ systematics are different; the earlier estimate for the
correction to $\lambda_L$ gives $\lambda_L = .160$ in this case, so one
should have expected a larger surface tension.

The one loop analytic estimate of the surface tension is
\begin{equation}
\int_0^{\phi_0} d\phi \sqrt{2 V_1 (\phi)} \, ,
\end{equation}
where $V_1$ is the one loop effective potential.  For $\lambda_L = 0.167$
the result is $.0254 g^4 T^3$ and for $\lambda_L = .160$ it
is $.028 g^4 T^3$.  One may also apply the same formula using the
two loop effective potential, though this is slightly inconsistent
since one is still using the tree level kinetic term without one loop
wave function corrections.  The results are $.080 g^4 T^3$ and 
$.086 g^4 T^3$ respectively.  The large difference between the one and
two loop perturbative estimates reflects the $\phi^3$ dependence of
the surface tension.  Since the two loop perturbative value for $\phi$
and the lattice value are quite close, it is not surprising that the
two loop surface tension is quite close to the lattice value.  The lattice
$\phi$ is larger than the two loop value, but the lattice surface
tension is smaller; this may be the beginning of a trend of low
surface tensions found in \cite{KLRSresults}, though it is difficult to
say until the remaining $O(a)$ effects are accounted for.  A
lower surface tension is also 
the direction one would expect from including wave function corrections,
as discussed in \cite{Kripfganz}.

We conclude that the microcanonical technique is an efficient and
promising way of extracting all interesting thermodynamical properties
of the phase transition; it should be further pursued 
after a more careful accounting of $O(a)$ corrections has been made.

\section{Friction on the bubble wall}
\label{friction}

The previous section merely uses the Hamiltonian evolution of classical
Yang-Mills Higgs theory as a microcanonical Monte-Carlo algorithm for
thermodynamic investigation, but as stressed in Section 
\ref{defendclassical}, the classical Hamiltonian evolution should also
give information about dynamics, provided that the physics involved is
infrared dominated.  This includes two previously elusive phenomena,
friction on the bubble wall from infrared bosons and the motion of
Chern-Simons number near a moving bubble wall.

\subsection{General discussion}

The velocity attained by a moving bubble wall during the cosmological
electroweak phase transition is one of the key ingredients for 
models and calculations of baryon number production.  Several authors
have considered the problem
\cite{Turok,Ignatius,Dine,Liu,Khlebnikov,Arnold,Heckler,MP1,MP2,Laine},
and quite a bit is known.  Two effects prevent the bubble wall
from ``running away'' and establish its terminal velocity;
frictive effects arising from the departure from equilibrium 
of massive species due to the motion of the wall, and 
hydrodynamic effects arising from the liberation of latent heat.
The hydrodynamic properties of the plasma are dominated by thermal
energy particles, which hold almost all of the energy and momentum of
the plasma, so these effects cannot be studied by classical techniques.
Fortunately, except perhaps for energy transfer across the bubble
wall by ballistic leptons, the hydrodynamics are well under control
\cite{Ignatius,Heckler,MP2,Laine}.  

The friction from massive particles depends on a high power of the 
particle mass, and so the only important particles for consideration 
are the top quarks and the bosons of the Yang-Mills Higgs system.  Friction
from top quarks cannot be calculated in the classical theory, but it
arises at a higher parametric order than the friction from bosons,
where the leading parametric contribution is infrared dominated and 
should be reproduced in the classical simulations.

To see this, we begin by stating what we mean by the friction.  We will
only be interested in this paper with the friction in the case that
the departure from equilibrium is small, so that a fluctuation
dissipation formula can relate it to measurable equilibrium correlators
\cite{Khlebnikov}.  In this case we define the friction coefficient
as
\begin{equation}
\eta \equiv \lim_{P \rightarrow 0} \frac{P}{v_w} \, ,
\end{equation}
where $v_w$ is the average velocity of a planar bubble wall when the 
pressure difference between phases is $P$.  As we will discuss below, 
this friction coefficient is related to the diffusion constant for
the random motion of the equilibrium interface, and converting 
from lattice units to thermal units establishes the parametric behavior
of the classical $\eta$ as $\eta_{cl} \propto g^6 T^4$.

The friction coefficient for the quantum theory can be computed at 
lowest order in a loopwise expansion from a fluctuation-dissipation
theorem \cite{Khlebnikov}, and the answer is equivalent to the friction
from free scattering, thermal particles on the bubble wall \cite{Arnold},
which has been calculated in \cite{Dine,Liu}.  
The contribution from the 6 transverse $W$ boson modes is 
\begin{eqnarray}
\label{1loopfriction}
\eta & = & 6 \left[ \int_0^m \frac{E dE}{4 \pi^2}\frac{1}{e^{\beta E}-1}
2 E^2 \right. \\ & & \left.
+ \int_m^{\infty} \frac{E dE}{4 \pi^2} \frac{1}{e^{\beta E} - 1}
\left( 2 E^2 - m^2 - 2 E \sqrt{E^2 - m^2} \right) \right] 
\nonumber \\
& \simeq & 6 \left[ \frac{g^3}{32 \pi^2} T \phi^3 -
	\frac{g^4 \phi^4}{512 \pi^2} \left(\frac{7}{4} + 
\ln \frac{8 \pi T}{g \phi} - \gamma_E \right) \right] \, ,
\end{eqnarray}
where the first integral arises from particles which scatter from the
wall and the second is from particles which fly over the wall from each
side.  If we make the approximation $1/(\exp( \beta E ) - 1) \simeq 
(1/\beta E)$, which is precisely the classical approximation to the 
Bose-Einstein population factor, then the integrals give precisely
the first $O(g^3 \phi^3)$ term.  Hence we can understand the result
as a classical part, which (recalling that $\phi \simeq g T$) is
order $g^6 T^4$, plus a correction which is $O(g^8 \ln 1/g)$.
The most important point is that the result is totally dominated by
the infrared contribution, and extending the classical approximation for
the Bose distribution out to arbitrarily large momenta only makes
a parametrically suppressed error; hence the friction from bosons
is (at leading parametric order) an infrared, classical effect.  This means, 
of course, that the loopwise expansion will be unreliable, because
the infrared is strongly coupled; instead we should use classical
real time simulations to determine the friction.

For the fermions, Eq. (\ref{1loopfriction}) applies but with the 
Bose-Einstein statistics replaced by Fermi-Dirac statistics 
$1/(\exp(\beta E) + 1)$ (and of course the 6 replaced by a 12).
The resulting friction is
\begin{equation}
\eta \simeq 12 \left[ \frac{g_Y^4 \phi^4}{128 \pi^2} \left( \frac{7}{4} +
	\ln \frac{\sqrt{2} \pi T}{g_Y \phi} - \gamma_E \right) \right] \, ,
\end{equation}
which begins at $O(g^8 \ln 1/g)$.  Friction from top quarks is 
parametrically suppressed and will not 
appear in the classical theory, which is obvious because the classical
theory is purely bosonic.  The top quark contribution to friction 
may still be significant, simply because
it depends on $g_Y^4$ and $g_Y$ is numerically much larger than $g$.
The absence of a large infrared contribution should also make the top
quark friction computable, although the naive perturbative series must be
resummed into Boltzmann equations to account for on-shell near 
singularities \cite{Jeon} and may then require simplifying analytical
approximations to make the calculation feasible \cite{MP1,MP2}.

Fortunately, most of the top quark friction arises from the small departure
from equilibrium of thermal energy particles; because Fermi statistics
are well behaved in the infrared, the departure from equilibrium of
top quarks will have a small (parametrically suppressed) influence on
the infrared bosonic sector.  The two frictions should be additive 
with $O(\alpha)$ error, and an investigation of the friction in the
bosonic theory is well motivated.

\subsection{Fluctuation dissipation relation}

As mentioned above, the $O(g^6)$ contribution to the 
friction coefficient can be computed in the
limit of a small departure from equilibrium by a fluctuation dissipation
argument, as follows.  
Consider an infinite square tube of cross section $A$, filled with 
classical Yang-Mills Higgs plasma at the equilibrium temperature.  
Two semi-infinite regions of definite phase are
separated by a domain wall.  Because the system has translational 
invariance, for
times much greater than any thermalization time in the plasma the
position $x$ of this wall will diffuse,
\begin{equation}
\lim_{t \rightarrow \infty} \frac{ \langle (x(t) - x(0))^2 \rangle }
{ t } = D_x \, ,
\end{equation}
with $D_x$ the diffusion constant for $x$.
Hence, the probability that, starting from $x(0)$ at time $0$, one
will arrive at $x(t)$
at time $t$, which we call ${\cal P}(x(t),x(0),t)$, satisfies
\begin{equation}
\int dx(t) {\cal P}(x(t),x(0),t) (x(t)-x(0))^2 = |t| D_x \, .
\end{equation}

Now suppose that we exert a very small force $f(x) = -V'(x)$ 
on the bubble wall.  We choose $V'$ to be constant in a large
neighborhood of the origin, although $V$ eventually turns up so that it
is bounded from below and goes to infinity at $x \rightarrow \pm \infty$.
At leading order the hopping probability will be modified by an offset,
\begin{equation}
\int dx(t) {\cal P}_f(x(t),x(0),t) (x(t)-x(0)) = 
\langle x(t) - x(0) \rangle = C ft = v_w(f) t
\end{equation}
with $C$ a constant to be determined.  Also, the equilibrium
probability distribution of
$x$ will be multiplied by a Boltzmann factor $\exp(-V/T) = \exp(fx/T)
\simeq 1+fx/T + (fx)^2/2T^2$.  Starting with a thermal distribution of
wall positions at time $t=0$, the probability that the wall should be
at position $x(t)=0$ at time $t$, relative to its probability for
starting there, is
\begin{equation}
1 = \int dx(0) {\cal P}_f(0,x(0),t) \left( 1 + \frac{fx(0)}{T} +
 \frac{x(0)^2 f^2}{2T^2} \right) = 1 - \frac{Cf^2t}{T} + \frac{D_x
f^2 t}{2 T^2} \, ,
\end{equation}
and hence
\begin{equation}
C = \frac{ D_x }{2T} \, .
\end{equation}

The pressure on the wall is $P = f/A$, so the friction coefficient is 
\begin{equation}
\eta = \frac{P}{v_w} = \frac{P}{fC} = \frac{2T}{D_x A} \, .
\end{equation}
In physical units,
\begin{equation}
D_{x \; {\rm phys}} = \frac{4 D_{x \; {\rm latt}}}{g^2 \beta_L T}
\qquad {\rm and} \qquad 
A_{\rm phys} = \frac{16 A_{\rm latt}} {g^4 \beta_L^2 T^2} \, ,
\end{equation}
so $\eta \propto g^6 T^4$, as stated above.
To compute the bosonic contribution to the friction coefficient at
leading parametric order it is then only necessary to
determine $D_x A$ for a bubble wall in a classical simulation.

It is necessary to make a few changes to the above ideas to implement
them in a realistic simulation.  The simulation occurs in a box with
symmetric boundary conditions; there are two bubble 
surfaces, and there is a
slight mutual interaction between them, depending on their separation;
for instance, the motion of one wall releases latent heat, which
slightly changes the temperature and thereby induces a response from
the other wall;
so even if each wall's position can be described by a single
coordinate, the motion of the two coordinates will be interdependent.
The interdependences appear in the motion of the difference of the
wall positions, but translational invariance ensures diffusive motion
of the average of the wall coordinates.  The above arguments apply,
except that the force $f$ must be the total force on both walls, and
$D$ should be computed for the average coordinate.  The
same final expression holds, except that $A$ now represents the sum of
the areas of the two walls, twice the cross section of the box.
In terms of the cross-section of the box and the diffusion rate of the
average coordinate $D_{\rm av}$, 
\begin{equation}
\eta = \frac{T}{D_{\rm av} A} \, .
\end{equation}

\subsection{Numerical results}

We determine the wall position as follows.  Every few lattice lengths
in time, we average $\Phi^{\dagger} \Phi$ over the short directions of
a long rectangular lattice, producing a c-number function of the long
direction.  this is smoothed with a Gaussian broad enough to make the 
phases very clearly distinct but well narrower than the wall.  For
$\beta_L = 8$ and $\lambda_L = 0.2$ we found $\sigma = 3$ lattice
units sufficient, but for larger $\lambda_L$ where the phase transition
is weaker the smoothing would need to be stronger (and the cross-section
of the box would need to be wider).  We set a threshold at some fraction
of the way between the average broken phase value and the average 
symmetric phase value of $\Phi^{\dagger} \Phi$ (typically $30 \%$) and,
starting at the point of minimum $\Phi^{\dagger} \Phi$, move out in
either direction and identify the wall as the first point where the
threshold is exceeded.  This works well because, as noted earlier, the
fluctuations in $\Phi^{\dagger} \Phi$ are much smaller in the symmetric
phase than in the broken phase.

\begin{table}
 \begin{tabular}{|c|c|c|c|c|}\hline
lattice size & $\beta_L$ & $t$ (lattice units) & $D_{\rm av}$ (lattice units)
& $\eta$ \\ \hline
$16^2 \times 192$ & 8 & 1100 & 
$ 1.21 \pm 0.24 $ & $.026 \pm .005 g^6 T^4$  \\ \hline
$32^2 \times 192$ & 8 & 640 & 
$ .25 \pm 0.05 $ & $.031 \pm .006 g^6 T^4$  \\ \hline
$36^2 \times 144$ & 6 & 674 & 
$ .095 \pm 0.015 $ & $.027 \pm .004 g^6 T^4$  \\ \hline
\end{tabular}
\caption{\label{frictiontable} Data from computing the friction on the
bubble wall.  Three runs on three lattice sizes and at two lattice
coarsenesses are within error.}
\end{table}

\begin{figure}
\centerline{\psfig{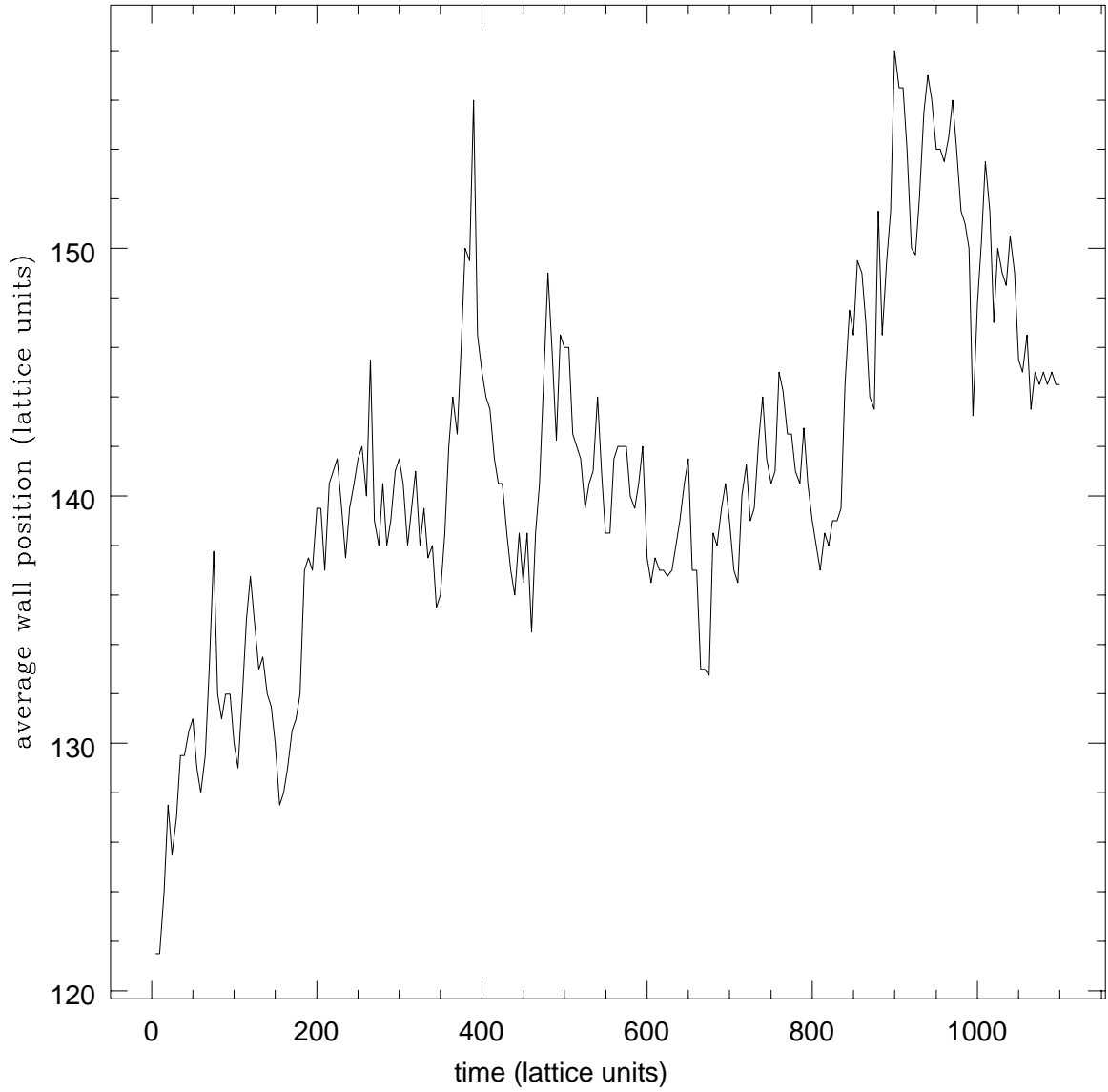}}
\caption{\label{wallpos} Average of positions of bubble walls as a
function of time for a $16 \times 16 \times 192$ box at equilibrium
with $\beta_L \simeq 8$.  The motion should be Brownian.}
\end{figure}

\begin{figure}
\centerline{\psfig{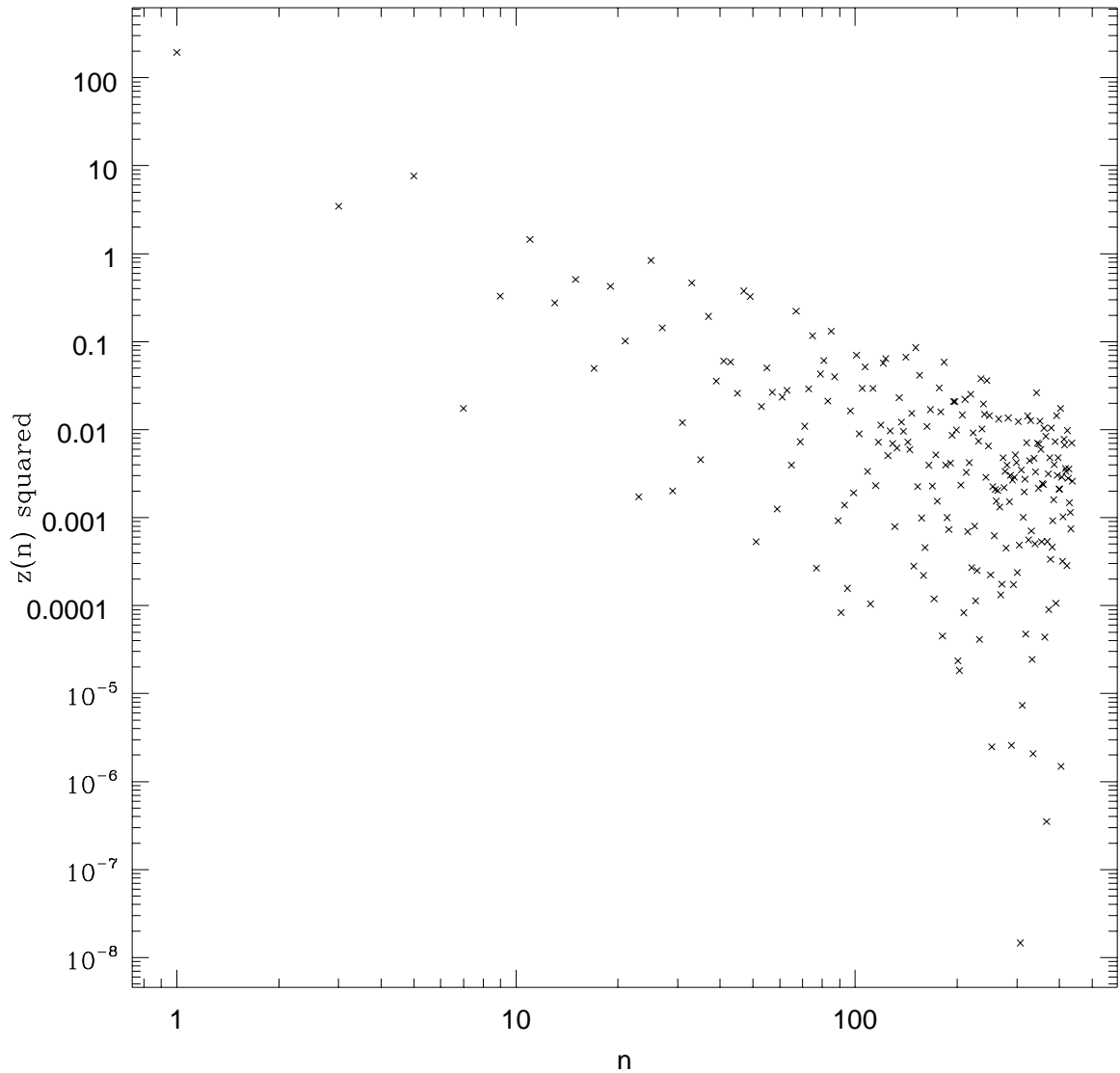}}
\caption{ \label{wallposft} Power spectrum for the previous plot,
obtained by sine transformation.  The wide scatter is expected for the
log of the square of a Gaussian distributed quantity.  If the motion
is Brownian then the points should be uncorrelated and should depend
on $n$ as $1/n^2$.}
\end{figure}

In Figure \ref{wallpos} we plot the average of the positions of the 
two walls in a $16 \times 16 \times 192$ box.  The data are taken every
5 lattice lengths for a period of 1100 lattice lengths.  We extract
a diffusion constant from this data by the method discussed in 
Appendix A, which uses the fact that the coefficients of a sine 
transform of a Brownian process are independent and Gaussian, with
variance $\propto 1/k^2$.  The power spectrum is plotted in 
Figure \ref{wallposft},
which clearly shows this powerlaw behavior.  Results for this run,
for a 640 lattice length run in a $32 \times 32 \times 192$ box,
and for a 674 lattice length run in a $36 \times 36 \times 144$ box
at $\beta_L = 6$, are presented in Table \ref{frictiontable}.  In each
case several (order 10) of the lowest frequency Fourier modes were
removed, as discussed below, and around 100 of the next lowest modes
were used; all higher frequency modes, which are more polluted by 
small errors in the wall finding algorithm and which may contain 
physics on time scales too short for the diffusion description to be 
valid, are also excluded.  We checked that $ 50 \% $ changes to 
these cuts did not
significantly affect the results.  The error bars reflect the quality of
the fit, see Appendix A, and
are typically twice the change resulting from a $50 \%$ change in
the cuts.  The runs are mutually consistent, which is a good test since
they were performed on lattices of different sizes and coarsenesses.
For comparison, the friction expected from the one loop (free scattering)
estimate, from transverse $W$ bosons alone, and using the
lattice value of $\phi=1.54 g$, is $ .069 g^6 T^4$, 
larger by a factor of about 2.5.  One expected the free particle
estimate to exaggerate the friction, 
because scatterings will thermalize the particles as they reflect from
the wall, reducing the friction; but the amplitude of this reduction
could not previously be calculated reliably.

We should mention a serious note of caution in establishing these
results and in measuring the friction on a bubble wall in a classical
simulation.  This has to do with hydrodynamics.  As noted earlier, 
the liberation of latent heat by a moving bubble wall raises the temperature
of the plasma locally, and the heat is redistributed by 
the plasma.  The temperature at the bubble wall is generically
elevated with respect to the temperature at the time the bubble
nucleated, slowing the wall in a way which cannot strictly be 
said to be frictional.  In the quantum system heat is transported by
hydrodynamic waves which travel at the speed of sound $v_s = 1/\sqrt{3}$,
and for an isolated, spherical bubble, the dissipation of heat into
the surrounding medium is efficient if the wall velocity is small.  

Latent heat is also liberated in the classical lattice system, but
the specifics of the hydrodynamics are cutoff dependent and qualitatively
different.  For instance, in physical units the heat capacity of the
classical plasma increases with inverse lattice spacing as $a^{-3}$, 
while the latent heat depends on $a^{-1}$; so the relation between the
latent heat and the energy drop between equilibrium and nucleation
depends on $a$.  Worse, heat is not conveyed on the lattice by bulk
hydrodynamic flow; it travels diffusively.  The reason is that the lattice 
does not respect momentum conservation; while momentum is approximately 
conserved for interactions between infrared modes, many ultraviolet
modes obey dispersion relations in which
adding momentum accelerates the mode in the opposite direction
of the applied momentum.  Hence, scatterings cause momentum to be
lost into the lattice, and bulk flows tend to come to a stop; on 
large length scales heat dissipates rather than flowing hydrodynamically.
We have verified this numerically by heating the lattice system
inhomogeneously so as to excite the fundamental standing wave;
the amplitude of the standing wave decays rather than oscillates.

The problem lies in preventing the hydrodynamic response of the classical
fluid from contributing to the observed friction.  In the case of the
diffusion rate mentioned above, this becomes a problem for the very
low frequency Fourier modes.  The latent heat liberated by the motion
of the wall tends to sit on top of the wall, diffusing away only slowly,
so there is a very long term ``memory effect'' inducing negative
correlations in the wall motion.  In Figure \ref{WallPower} we see 
that the most infrared sine transform coefficients 
do indeed fall off the power law curve;
this is why it is necessary to drop the most infrared coefficients
from the analysis of the diffusion constant.  Between the low frequency
coefficients which must be removed and 
the high frequency coefficients which do not show diffusive
behavior there is a sizeable fiducial range, as we established by moving
the cuts in and out on either side by $50 \%$ and finding little difference
to the determined diffusion rate, as discussed above.
The cut for $ \beta_L = 6 $ had to be sharper because
the heat capacity is smaller compared to the latent heat in that
case.

Another way of measuring the infrared contribution to the bubble wall
friction is to supercool the system in the symmetric phase and 
produce a small region of broken phase by some means, for instance
by briefly making $m_{H0}^2$ more negative in a narrow region, or 
by starting with a mixed phase configuration and applying the
cooling very abruptly.  One then directly measures the velocity at which
the bubble walls move and sweep up the symmetric phase, averaging over
several repitions with different initial conditions 
to determine and improve statistics.
This technique has the advantage that it
does not rely on a linear approximation, which may not be very good
if the departure from equilibrium is large; but  the latent heat 
poses a problem for this technique.  For the lattice
coarseness and value of $\lambda_L$ used in this paper, the temperature rise
due to the liberation of latent heat is comparable to the supercooling,
as can be seen from how much the temperature jumps on changing phase 
in the hysteresis plots, see Figure \ref{OPjump}.  Another 
problem is that, if the departure from linear response really is 
important, then the frictive pressure may not be linear in $v_w$, and
it becomes more difficult to add the effects of the top quark and 
the $W$ boson.

We end this section by commenting on possible ways around the 
hydrodynamics problem.  One way is to work on a much finer lattice,
so that the heat capacity is large enough to soak up the latent heat
without much effect.  The obvious disadvantage is that this technique
is numerically expensive.  Alternately, one could add extra fundamental
representation, massive scalars to the theory to serve as an additional
heat reservoir.  If their masses are chosen large enough then they
will have little effect on the infrared physics.  They would also 
tend to raise the Debye screening mass, which might be desirable.
As long as they had no coupling to the Higgs boson, one would not
expect them to contribute to the friction, except by their interactions
with the other particles.  A final possible solution is to turn on a
very weak Langevin noise and damping term.  The technology for doing
this in a gauge invariant way was worked out in \cite{Krasnitz}.  As a
thermalization algorithm the technique is non-optimal because it
thermalizes the infrared modes much more slowly than
the ultraviolet ones; but for this application that is a blessing, since
one wants to absorb the heat going into the ultraviolet excitations 
without interfering with the correct infrared dynamics.  The main 
disadvantage is that the Langevin term makes the evolution canonical,
rather than microcanonical, so the two phase configuration would
be unstable; but in a large enough box it would take a
long time for the walls to diffuse into each other.  We will not
pursue any of these techniques here.

\section{Chern-Simons number motion on the bubble wall}

\label{NCSsection}

Another question which the real time technique can answer is how the
system responds to a chemical potential or other driving force for the
motion of Chern-Simons number $N_{CS}$.  This is directly relevant to
the study of baryon number violation because the axial anomaly relates
the motion of baryon number to the motion of $N_{CS}$.  In fact the
classical technique was originally proposed to investigate this 
question \cite{GrigRub}, and since then it has been used to investigate
$N_{CS}$ violation in Yang-Mills theory
\cite{AmbKras,Moore1,Moore2}, in the symmetric phase of Yang-Mills Higgs
theory \cite{Amb1,Ambjornetal}, and in the broken phase of 
Yang-Mills Higgs theory \cite{newguys,Krasprivate}.
It has yet to be applied to perhaps the most relevant problem, which
is the rate of $N_{CS}$ motion in the out of equilibrium environment
of a moving bubble wall during the phase transition.  The study of
mechanisms for baryon number violation at the electroweak phase
transition is the study of how the out of equilibrium phase transition
physics can induce in the infrared bosonic effective theory $CP$ violating
operators.  To convert this information into a baryon number abundance
we must understand how $N_{CS}$ responds to these operators.

Turok and Zadrozny \cite{TurokZadrozny} and
McLerran et. al. \cite{MSTV} take the fermions to be in equilibrium and
integrate them out, and then try to investigate how the resulting 
$CP$ violating operators will influence the out of equilibrium 
decay of gauge field configurations as they are swept onto the
wall; but for lack of quantitative tools they were forced to make only
qualitative or parametric estimates (see also \cite{NTrev}).
More recently work has focused
on how the transport of fermions can carry $CP$ 
violation away from the bubble wall \cite{CKN}, \cite{JPT},
%
%
into the symmetric phase. One of the main motivations for examining
the transport mechanism was the worry that 
baryon number violation on the wall would be
strongly suppressed except for at the
very leading edge \cite{Dinearg}, \cite{CKN},
\cite{DineThomas}, by the turning on of the sphaleron mass.
Thus a chemical potential inside the wall would be ineffective.  But 
it remains to be shown whether this picture, based on viewing the
baryon number violation as a completely local, quasiequilibrium process, 
is correct, or whether decaying gauge field configurations
swept up onto a bubble wall can still respond to $CP$ violating effects
on the wall to produce $N_{CS}$ change.  In this chapter we will 
focus on this question. 

To answer it we need to apply a chemical
potential for $N_{CS}$ on the lattice, as otherwise we cannot create
a $CP$ violating bias\footnote{Another possibility is to put $CP$ violating
high dimension operators, such as $\Phi^{\dagger} \Phi E \cdot B$,
in the action; although such terms cannot bias $N_{CS}$ in equilibrium,
the shift in $\Phi^{\dagger} \Phi$ during the phase transition will bias
topology change.  However, implementing such operators makes 
evolving the system much more complicated, because
the update rule becomes nondiagonal in the natural basis of degrees of
freedom; also any nonrenormalizeable operator has much more influence on
the ultraviolet behavior of the lattice system than on the infrared,
which is potentially dangerous.}.  The technology for doing this has
recently been worked out for Yang-Mills theory \cite{Moore1},
and it is straightforward to extend it to Yang-Mills Higgs theory.
We briefly review the method in Appendix A, and extend it to 
Yang-Mills Higgs theory.  For a complete exposition see \cite{Moore1,Moore2}.

The only problem in applying a chemical potential for 
 $N_{CS}$ in Yang-Mills Higgs theory
is that the rate of $N_{CS}$ motion (or the rate of $N_{CS}$ diffusion
when no chemical potential is applied) is slightly contaminated by
ultraviolet lattice artefacts; in the broken phase these effects 
contribute about $1/10$ of the symmetric phase rate to the response.  We give
evidence that this rate is an ultraviolet lattice artefact 
in Appendix A.  The consequence
of this problem is that we must observe a rate of $N_{CS}$ motion
well above 1/10 the symmetric phase response rate before we know that
it corresponds to genuine infrared processes and not to spurious 
ultraviolet physics.

With this in mind, we set out to investigate how $\dot{N}_{CS}$
interpolates between the symmetric phase value and
the broken phase value across a bubble wall, when there is a spatially
uniform chemical potential for $N_{CS}$ in both phases.  We do this 
first in equilibrium.  We take a mixed phase configuration with about
equal volumes of broken and symmetric phases, in a $16 \times 16 \times 192$
box at the equilibrium temperature and the values of $\lambda_L=0.2,$
$m_{H0}=-.3223$ generally used in this paper.  We compute $\dot{N}_{CS}$
at each lattice point and bin it according to its distance from the
bubble wall.  This means that we find the bubble wall surfaces by
the algorithm described in Appendix B, and for each point in the plasma
we find the minimum vertical distance to a bubble wall, and add
$\dot{N}_{CS}$ to a bin corresponding to that distance, considered
positive if the point is on the broken phase side of the wall and negative 
if it is on the symmetric phase side.  Binning this way accounts for the
fact that the wall is an uneven surface.  We also bin $\Phi^{\dagger}
\Phi$, unsmoothed, using the same rule; this gives a bubble wall
profile.

\begin{figure}
\centerline{\psfig{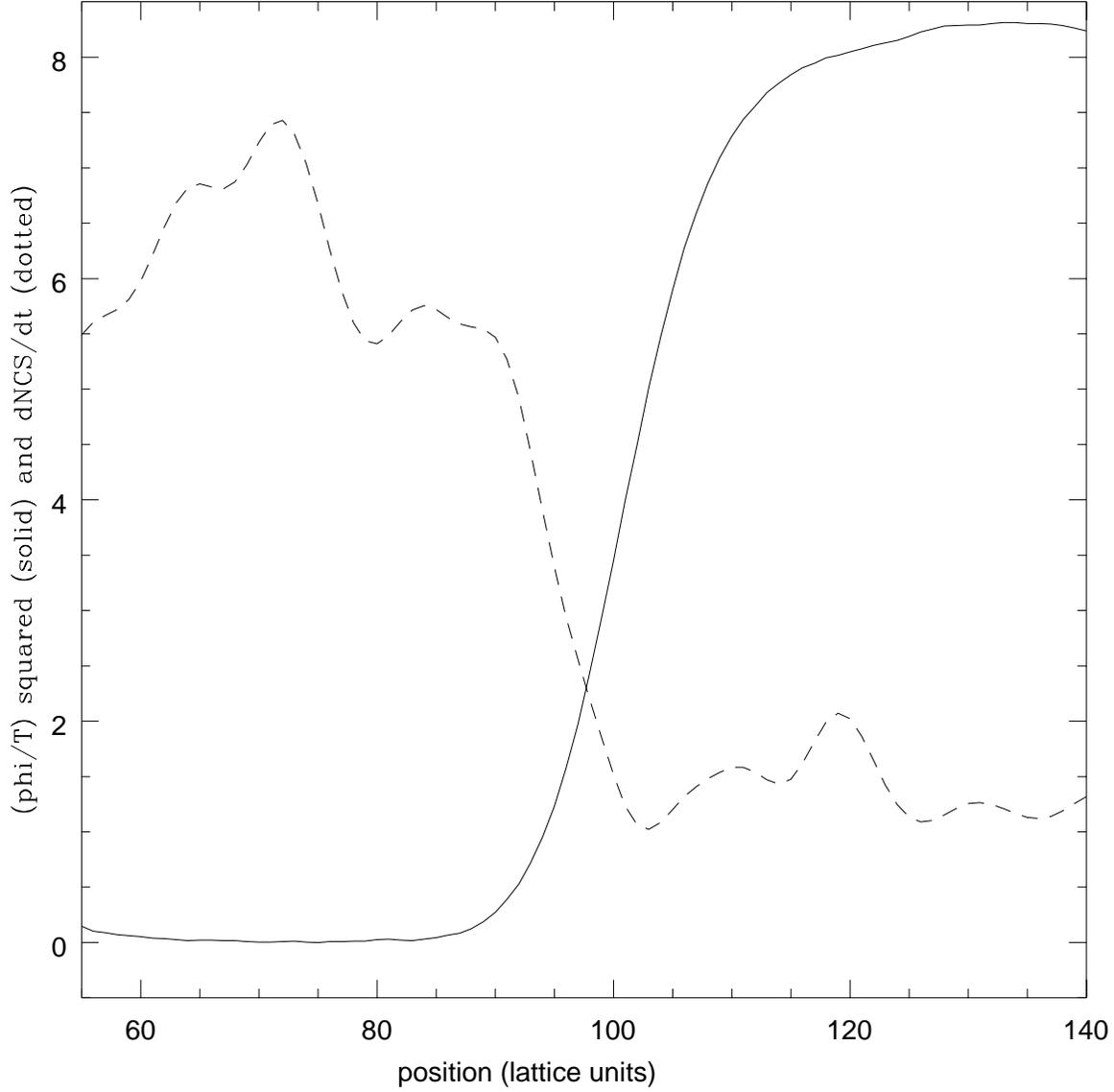}}
\caption{\label{atrest} Wall shape in equilibrium 
(solid line).  The vertical
axis is $\Phi^\dagger \Phi \beta_L^2$ in lattice units, 
equal to $4 (\phi^2/g^2 T^2)$ in the continuum. 
The symmetric phase value has been subtracted.
The  horizontal 
axis is in lattice units, which equal $4/(\beta_L g^2T) \simeq 1.2/T$ in
physical units; the zeropoint of the axis is arbitrary.
The dashed line shows $dN_{CS}/dt$ in reponse to a constant
chemical potential, as a function of position relative to
 this wall.  The vertical scale for $\dot{N}_{CS}$ is 
arbitrary. The rate falls off sharply 
inside the wall.}
\end{figure}

The results for the equilibrium case with a space independent chemical 
potential are presented in Figure \ref{atrest}.  
The $\dot{N}_{CS}$ results have been smoothed with a Gaussian envelope
of width $\sigma = 3$ lattice units to eliminate white noise fluctuations,
while the $\Phi^{\dagger} \Phi$ plot is completely unsmoothed; its
smoothness arises from averaging over a long time (approximately 4000
lattice lengths) and over the area of the bubble walls.
What we see in these plots is that, in equilibrium, the response to a
chemical potential falls off abruptly just inside the bubble wall,
and sphaleron events are suppressed in most of the interior of the
wall, as well as behind it.

Next we consider the out of equilibrium case.  We produce a series of
initial conditions by evolving a mixed phase, equilibrium configuration
through a series of short ($t=10$ lattice unit) Hamiltonian
trajectories.  The starting configuration contains mainly symmetric phase, 
so the broken phase can expand for some distance before the walls meet.
For each initial condition, we evolve it using the ``local thermostat'' 
discussed in Section \ref{Numerics} to drive the temperature to a point
midway between equilibrium and the spinodal 
point; we also turn on a global chemical potential for $N_{CS}$.  After
waiting about $t=20$ lattice units for the system to achieve a steady state,
we begin to record $\dot{N}_{CS}$ and $\Phi^{\dagger} \Phi$, binning
according to proximity to the bubble wall.  We discontinue the
evolution well before the bubble walls meet.

\begin{figure}
\centerline{\psfig{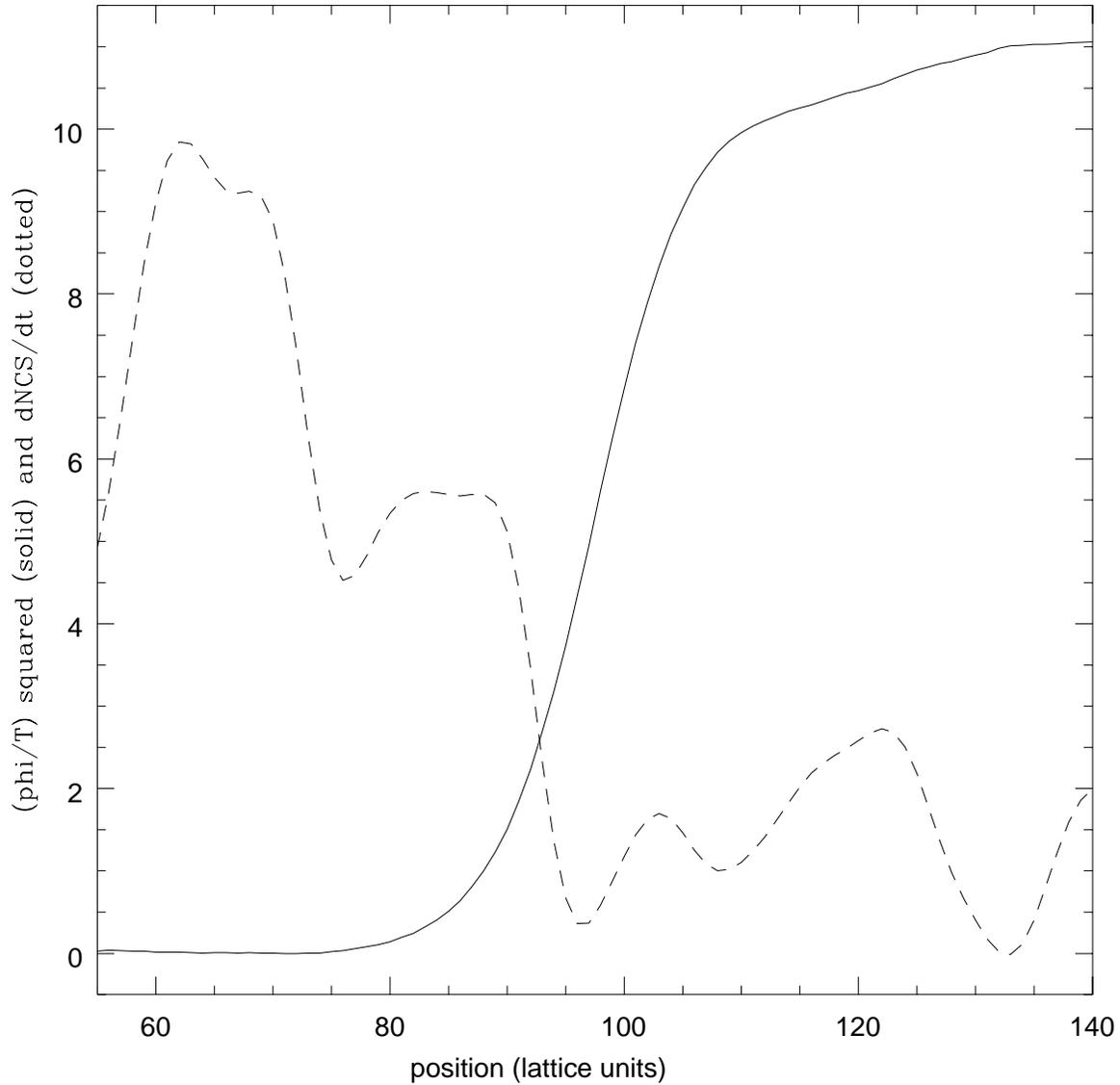}}
\caption{\label{moving} Wall shape out of equilibrium, below the 
phase transition temperature (solid line). 
The wall is in this case moving with
a speed of $v \sim 0.3$.
The dashed line shows $dN_{CS}/dt$  in reponse to a constant
chemical potential, as a function of position relative to
this wall.  Again the
rate falls off sharply inside the wall. Axes as in previous Figure.}
\end{figure}

The wall shape for the out of equilibrium case, and 
the motion of $N_{CS}$ in this case, with a constant chemical 
potential, are presented in
Figure \ref{moving}.
The conclusion is similar to the case
of the wall at rest; baryon number shuts off some short distance into
the bubble wall.

We also investigate the case where the chemical potential exists only
on the bubble wall.  We could do this by adding a nonrenormalizeable
operator $\Phi^{\dagger} \Phi E \cdot B$ to the action, which 
closely resembles an operator which would arise in the two Higgs
doublet model.  However there are serious technical difficulties with
adding such a term; it significantly modifies the ultraviolet
dynamics, and it renders the update rule nondiagonal.  Instead we will
``mock up'' the effect of this term.  First, we measure the averaged
dependence of $\Phi^{\dagger} \Phi$ on distance from the bubble
wall, which is shown in Figure \ref{moving}.  
Now for each point in the plasma, we find
the vertical distance to the nearest bubble wall, and apply a chemical
potential at that point which is proportional to the derivative of 
the wall profile in Figure \ref{moving} at that value.  
The chemical potential is then only nonzero on
the wall, and in a way which mimics $d(\Phi^{\dagger} \Phi)/dz$,
which would be $\propto d(\Phi^{\dagger} \Phi)/dt$ for steady motion
of the bubble wall.

\begin{figure}
\centerline{\psfig{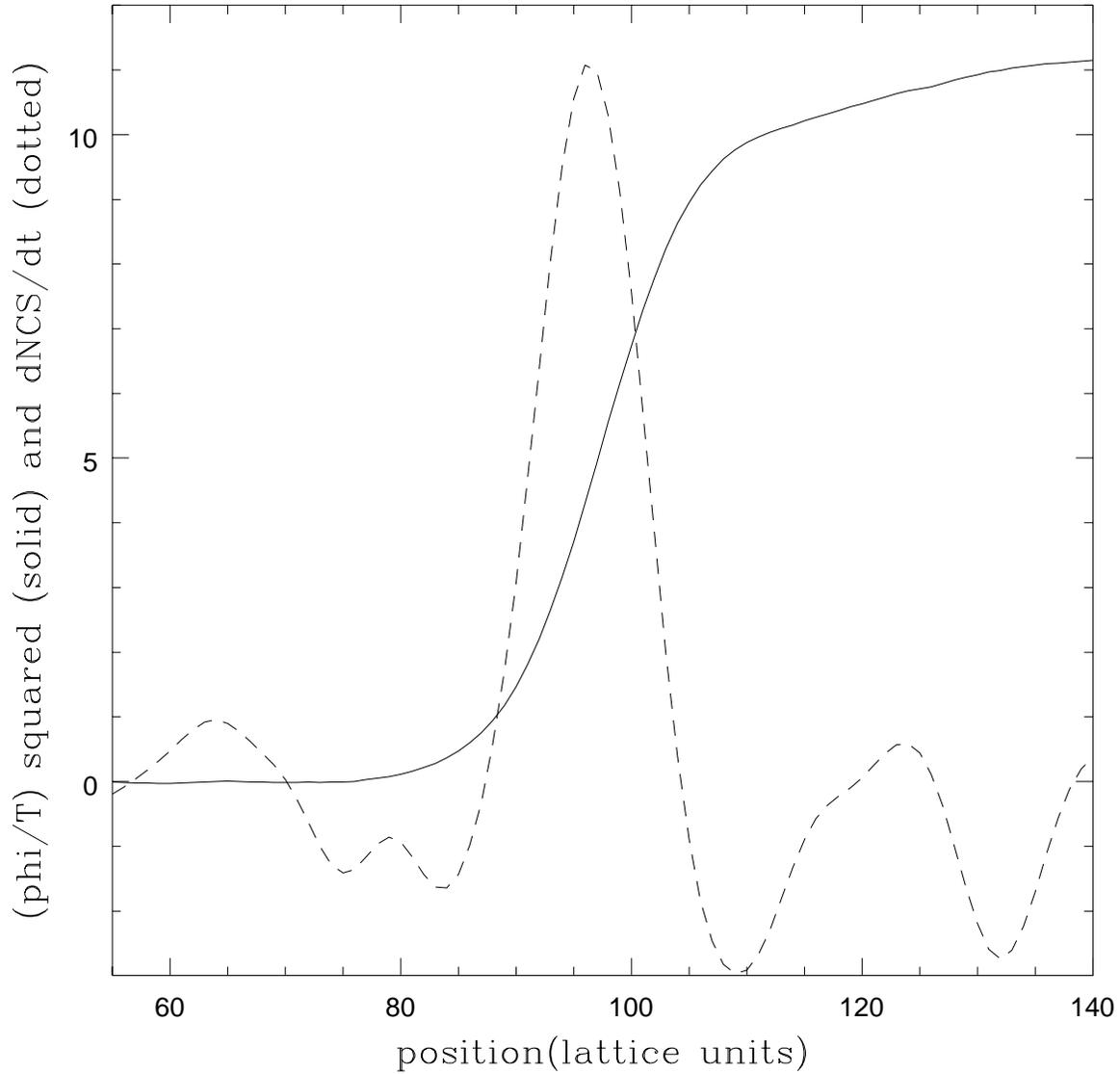}}
\caption{ \label{newfig} Bubble wall shape and $\dot{N}_{CS}$ when the
chemical potential for $N_{CS}$ was proportional to the gradient of
the wall.  The vertical axis for $\dot{N}_{CS}$ is arbitrary.
There is a spike in $\dot{N}_{CS}$ on the
wall, where the chemical potential was applied, 
and a pit on either side of the spike.}
\end{figure}

We have performed a series of runs with a chemical
potential which is only nonzero ``on the wall''.
The bubble wall shape and the average of
$\dot{N}_{CS}$ for each run are plotted in Figure \ref{newfig}.
The figure shows that $N_{CS}$ is generated 
in the interior of the bubble wall,
where the chemical potential is being applied, but it is destroyed in
a region immediately to either side.  This is not a fluctuation due
to insufficient statistics; if the data set is split in four, each 
quarter shows
the same morphology.  The rate of baryon number generation within
the hump itself is, in the dimensionless units discussed in Appendix A,
\begin{equation}
\frac{ (\beta_L \pi)^4 \int_{\rm hump} \dot{N}_{CS} d^3 x dt}
	{ \int \beta_L \mu(x,t) d^3 x dt } 
	= 0.225 \, .
\end{equation}
However, when one integrates over the full simulation volume, including
the regions to either side of the bubble wall where $\dot{N}_{CS}$ is
negative, one finds
\begin{equation}
\frac{ (\beta_L \pi)^4 \int \dot{N}_{CS} d^3 x dt}
	{ \int \beta_L \mu(x,t) d^3 x dt } 
	= 0.091 \pm 0.025 \, .
\end{equation}
The error bar on the latter number is estimated assuming the error
is from diffusion of $N_{CS}$ in the symmetric phase volume in the
simulation, and is consistent with the error bar from statistics between
runs.  the final result is larger than the expected 
``UV artefact rate'' of $\kappa_{\rm UV \; artefact}
\simeq 0.055$ (see Appendix A), so one might conclude that there is
a net infrared generation of $N_{CS}$ from the chemical potential; however
the statistical significance is not very good, and the generation is about
an order of magnitude less efficient than in the symmetric phase.

Why is there a bump with a dip on either side, and how can the system
respond to a chemical potential on the wall?  
Recall why a chemical potential should not induce
baryon number violation in the broken phase.  If a chemical potential
is turned on briefly, it will push the gauge fields in the direction
which generates baryon number, and, briefly, $N_{CS}$ will be generated.
However, the fields will respond elastically to this impulse--after
oscillating in the direction of greater $N_{CS}$ they will ``bounce
back'' to the previous value of $N_{CS}$.  The temporal response to
an impulse of chemical potential will then be a forward surge in $N_{CS}$
followed by a resotration to the original value.  (We have tested this
statement by tracking $N_{CS}$ after such an impulse.)  Now the gauge
fields also propagate, so if a chemical potential is applied in a
spatially nonuniform way, the relaxation to the former value of $N_{CS}$
may take place not where the chemical potential was applied, but
nearby.  This would explain the generation of $N_{CS}$ on the wall and
the destruction of $N_{CS}$ immediately to each side.
In the case of the moving wall, some of the ``excited'' gauge
field modes relax inside the symmetric phase, because they 
propagate there.  Once
in the symmetric phase, their behavior can be different--without a Higgs
condensate present, the most infrared modes may no longer be oscillatory
and will not relax as much as they would have, so the deficit in 
$\dot{N}_{CS}$ on the sides of the wall need not be as large as the 
production on the wall.  The infrared gauge fields themselves can serve
to transport the $CP$ violation on the bubble wall to the symmetric
phase, where it can lead to baryon number nonconservation.  However,
our data show that this is quite an inefficient mechanism.


In a previous draft of this paper we found a considerably larger effect
from chemical potential on the bubble wall.  However, in the evolutions
used for those results we were insufficiently careful to stop runs
well before the bubble walls met; sometimes they would begin to collide,
causing the bubble wall finding algorithm to go awry, and possibly leading
to the application of chemical potential inside the symmetric phase rather
than on the bubble wall surface.  For the runs presented here we were 
more careful; the new runs also represent about 3 times the cumulative
evolution time (totaling $t=14000$ lattice units).

\section{Conclusion}

\label{conclusion}

We have reviewed the fact that the thermodynamics of
quantum Yang-Mills Higgs theory coincides, in the approximation of
dimensional reduction, with the thermodynamics of the classical
theory, and shown how this can turn the real time evolution of
the classical theory into a powerful microcanonical Monte-Carlo
algorithm.  We used the algorithm, for Higgs potential parameters
corresponding to a tree level vacuum Higgs mass of $\simeq 50$GeV,
to investigate the phase diagram, including both metastable branches,
and to find the equilibrium temperature.  We also developed and 
applied technology for finding the equilibrium surface tension
by microcanonical techniques.

Further, we have strengthened the argument that the dynamics, as
well as the thermodynamics, of the infrared can be approximated
with classical field theory; the errors should be parametrically
suppressed (order $\alpha_W$) for phenomena which are truly
infrared dominated.
This is a consequence of the weakly coupled nature of the theory;
the physics can only become nonperturbative by becoming 
classical.  We were thus able to compute to leading parametric 
order the friction from bosons on a moving bubble wall in the near
equilibrium limit, and to investigate the physics of baryon number
violation out of equilibrium in the presence of a moving bubble wall.
We have found that the friction from infrared bosons is 
smaller than that from transverse $W$ bosons in the free scattering limit
by a factor of 2 -- 3, and baryon number violating processes proceed some
distance into the bubble wall but stop well short of the broken phase.

The classical method opens all thermodynamic and dynamical properties
which are dominated by the infrared bosonic sector to reasonably
accurate calculation.  To pursue the thermodynamics with good accuracy
it will be necessary either to apply a great deal of computer
time, or to make a thorough investigation of linear in $a$ corrections.
To improve the accuracy and reliability of the dynamical calculations
it will be useful in addition 
to find some way to properly represent hydrodynamic
effects on the lattice.  It may also be possible to integrate out
the $A_0$ field, so the large Debye mass limit is obtained without
necessitating very fine lattices.  

We are hopeful that such 
improvements in accuracy can be achieved, making it practical to use
the techniques developed here at a wide range of Higgs
masses and for interesting extensions of the Minimal Standard Model,
including a second light Higgs doublet and a light stop squark.

\centerline{Acknowledgements}

We would like to express our gratitude to Mark Alford, Peter Lepage, 
Ron Horgan, Chris Barnes, and Alex Krasnitz for useful 
conversations.  GM would like to thank DAMTP, Cambridge for
hospitality.  GM was supported under NSF contract PHY90-21984, and
NT was supported by PPARC, UK, and a startup grant from Cambridge
University.  The code used in this paper is
written in $c$, and is reasonably user friendly and well documented; it
is free on request to GM.

\appendix

\section{$N_{CS}$ in Yang-Mills Higgs theory}

Two quantities can be used to characterize baryon number violation in
classical Yang-Mills Higgs theory; the diffusion rate of $N_{CS}$,
and $\dot{N}_{CS}$ when there is a chemical potential for $N_{CS}$
added to the Hamiltonian.  In the large volume, long time limit these
quantities can be characterized by two rates,
\begin{equation}
\Gamma_d \equiv \lim_{t \rightarrow \infty} \frac{ \langle (N_{CS}(t)
- N_{CS}(0))^2 \rangle}{Vt} \quad {\rm and} \quad
\Gamma_\mu \equiv \lim_{\mu \rightarrow 0} \frac{ \dot{N}_{CS} T}
{ \mu V t} \, ,
\end{equation}
which satisfy a fluctuation dissipation relation, 
$\Gamma_d = 2 \Gamma_{\mu}$, derived in \cite{KhlebShap}
and further discussed in \cite{RubShap,Moore1}.
It is also interesting to know $\dot{N}_{CS}$ for $\mu \gg T$;
if $\dot{N}_{CS}$ rises much faster than linearly in $\mu$ it 
indicates that the motion of $N_{CS}$ is obstructed
by a substantial free energy barrier; for Yang-Mills theory 
$\dot{N}_{CS}$ turns out to be very linear so there is no free energy 
barrier \cite{Moore1}.

The technology for computing $N_{CS}$ as a function of time in 
Yang-Mills Higgs theory, in the absence of a chemical potential,
was developed in \cite{Ambjornetal,AmbKras}; we will give only the
most cursory review here.
Instead this appendix will concentrate on how best to extract a diffusion 
constant from the diffusive trajectory of $N_{CS}$, and then on how to
extend the chemical potential method developed in \cite{Moore1} to
the case of Yang-Mills Higgs theory.  We will then investigate $N_{CS}$
motion in the symmetric and broken phases, using both techniques, and
present three pieces of evidence that the rate we establish in the
broken phase arises from spurious ultraviolet effects.

\subsection{Finding a diffusion rate}

Suppose we know some quantity $z$ as a function of time $t$ between
an initial time $0$ and a final time $t_f$.  We believe that it should
exhibit Brownian motion, and we want to determine the diffusion
constant $\Gamma$, defined as $\langle (z(t_1) - z(t_2))^2 \rangle
= \Gamma | t_1 - t_2 |$.  One method, used in \cite{AmbKras}, is
to compute 
\begin{equation}
{\rm lag} (t_1) \equiv \int_0^{t_f-t_1}  (z(t+t_1) - z(t))^2 dt
\end{equation}
for many values of $t_1$ and to fit the result to a straight line.  
While this method works, the fitting is complicated by the large 
correlations between ${\rm lag}(t)$ for different values of $t$.  

We advocate an alternate method in which one sine transforms $z$ and
makes a likelihood analysis of the transform coefficients.  The analysis
is simpler because the sine transform coefficients are independent;
and the likelihood analysis is probably optimal in the sense of
statistical power.

We begin by redefining $z(t)$ as $z(t) - z(0)$; then
\begin{equation}
\langle z(t_1) z(t_2) \rangle = \Gamma  \; {\rm min} (t_1,t_2) \, .
\end{equation}
Defining the sine transform coefficient
\begin{equation}
\tilde{z}_n = \int_0^{t_f} \frac{dt}{t_f} z(t) \sin \left(
\frac{n \pi t}{2 t_f} \right) \qquad n = 1,3,5,7 \ldots
\end{equation}
we quickly find
\begin{eqnarray}
\langle \tilde{z}_n \tilde{z}_m \rangle & = & \int_0^{t_f} \frac{dt_1}{t_f}
\int_0^{t_f} \frac{dt_2}{t_f} \sin \left( \frac{n\pi t_1}{2t_f} \right)
\sin \left( \frac{n \pi t_2}{2t_f} \right) \Gamma \; {\rm min}(t_1,t_2)
\\
& = & \frac{2 \Gamma t_f}{mn \pi^2} \delta_{mn}  \, .
\label{powerspect}
\end{eqnarray}
The sine transform coefficients are independent, with variance
$2 \Gamma t_f/(n\pi)^2$.  They will also be Gaussian distributed.
We can see that the sines used form a complete 
set by extending $z$ as an odd function
about the origin and an even function about $t_f$.

In the case of interest, $z$ will only be known at discrete values
of $t$, and the integrals above represent the corresponding sums.
$z$ will also often be contaminated by white noise--this was proven
in the case of Yang-Mills theory by Ambjorn and Krasnitz, by considering
the Abelian approximation to the ultraviolet modes of the theory
\cite{AmbKras}--which will introduce a constant times $\delta_{mn}$
into Eq. (\ref{powerspect}).  The value of this constant is uninteresting,
but its presence means that the ultraviolet coefficients do not 
carry information about $\Gamma$, which can only be determined with
finite precision.  Any other non-Brownian correction to the ultraviolet
physics, as we expect for instance in extracting the diffusion 
constant for the bubble wall motion considered in Section \ref{friction},
also makes the ultraviolet data useless.

The remaining question is how, given a set of Gaussian distributed
coefficients $\tilde{z}_n$ , $n=1,3,5\ldots $, satisfying
\begin{equation}
 \langle \tilde{z}_n^2 \rangle = \frac{A}{n^2} + B \, ,
\end{equation}
to extract $A$ and its errorbars.  If we assume a uniform prior for the
values of $A$ and $B$, then Bayes' theorem gives the likelihood
of values $A$ and $B$ as
\begin{equation}
{\cal P}(A,B) \propto
\prod_{n=1,3,\ldots} \frac{\exp( -\tilde{x}_n^2 / 2 \sigma_n^2)}
{\sqrt{2 \pi} \sigma_n} \, , \qquad \sigma_n^2 = \frac{A}{n^2} + B \, .
\end{equation}
The best value of $(A,B)$ is that which maximizes this likelihood
function.  If the likelihood function is sharply peaked and 
there is a neighborhood of the maximum which contains almost all
of the probability, and in which $\ln {\cal P}$
is well approximated by a quadratic form, then the likelihood will
be approximately Gaussian distributed with an error matrix
which is the inverse of the quadratic form.  If there is enough
data to make a strong determination of $A$ and $B$ then this is generally
the case.  This is how we determine best fits and errors for Brownian
processes in this paper, and it will allow us to find the $N_{CS}$
diffusion rate in either electroweak phase.

\subsection{Chemical potential method in Yang-Mills Higgs theory}

It is useful to be able to measure the response of $N_{CS}$
to a chemical potential.  This idea was first investigated for Yang-Mills
Higgs theory by Ambjorn et. al. \cite{Ambjornetal}, who encountered technical
difficulties; these were overcome for Yang-Mills theory in \cite{Moore1}.
Here we briefly extend the method developed there to Yang-Mills Higgs theory.

The first step is to
define an (adjoint valued vector) magnetic field $B$; the appropriate choice
is \cite{AmbKras}
\begin{equation}
B_i^a(x) = \frac{1}{8} \sum_{4 \Box_{x,jk}} \frac{1}{2} {\rm Tr}
-i \tau^a U_{\Box} + \frac{1}{8} \sum_{4 \Box_{x+i,jk}} \frac{1}{2}
{\rm Tr} -i U_i(x) \tau^a U^{\dagger}_i(x) U_{\Box}
\end{equation}
where each sum is over 4 plaquettes orthogonal to the $x,i$ link, $U_{\Box}$
is the product of links around that plaquette beginning and ending
on the link, and the two sums are for the basepoint
and endpoint of the link. The $U_i(x)$ and $U^{\dagger}_i(x)$ in the
second term parallel transport $U_{\Box}$ to the basepoint of the link.
$\dot{N}_{CS}$ is then defined as
\begin{equation}
2 \pi^2 \dot{N}_{CS} = \sum_{x,i} E_i^a B_i^a \equiv E \cdot B
\end{equation}
where we have displayed the definition of inner product for adjoint
vector fields.  These two definitions are already sufficient for tracking
the (diffusive) motion of $N_{CS}$ in the absence of a chemical potential.

Next we see how to apply a chemical potential for $N_{CS}$.  
First consider Yang-Mills theory.
On the lattice it is not necessarily true that
\begin{equation}
0=\nabla \cdot B(x) \equiv \sum_i ( B_i(x) - U^{\dagger}_i(x-i)
B_i(x-i) U_i(x-i) )
\end{equation}
which means that if, for some reason, the Gauss constraint were
not actually obeyed, $\dot{N}_{CS}$ would depend on the size of
the violation.  To understand this point, suppose that we knew an
orthonormal basis for adjoint vector fields which partitions
into basis elements which are linear combinations of the constraints
(which can be constructed by 
taking the gradients of adjoint valued scalar fields)
and basis elements with zero divergence.  Call the projection
of the electric field into the former subset $E^c$ ($c$ for constraint)
and the projection into the latter subset $E^*$.  The Gauss constraint
is the condition $E^c=0$, and the $E^*$ are the dynamical degrees
of freedom.  Similarly, $B$ partitions into $B^c$ and $B^*$, and
the above point is that $B^c \neq 0$, so
\begin{equation}
2 \pi^2 \dot{N}_{CS} = E^c \cdot B^c + E^* \cdot B^*
\end{equation}
would change value if for some reason one orthogonally departed from 
the constraint manifold.  

This is the source of problems for the implementation of
a chemical potential for $N_{CS}$, which modifies
the $E$ equation of motion by
\begin{equation}
\label{wrongweigh}
\delta \dot{E} = - \mu \frac{ \partial \dot{N}_{CS}}{\partial E}
= \frac{- \mu}{2 \pi^2} B \, .
\end{equation}
Since $B^c \neq 0$ this moves $E^c$ away from zero, which in turn changes the
value of $\dot{N}_{CS}$, producing wrong answers.  What has happened
is that we have added a new term to the Hamiltonian, with a part linear
in $E^c$, so the constraint is no longer first class.  The evolution
then departs from the constraint manifold.
But if we redefine $\dot{N}_{CS}$ as
\begin{equation}
2 \pi^2 \dot{N}_{CS} = E^* \cdot B^* \, ,
\end{equation}
which coincides with the previous definition on the constraint manifold,
then the constraint will
again be first class and the evolution will not violate Gauss'
law.  The modification to the $E$ field equation of motion becomes
\begin{equation}
\delta \dot{E}^* = - \frac{\mu}{2 \pi^2} B^* \, , \qquad
\delta \dot{E}^c = 0 \, ,
\end{equation}
which preserves the Gauss constraints.  And since we never leave the 
constraint surface, we can still compute $\dot{N}_{CS}$ using $E \cdot B$.

The easiest way to implement this addition to the 
equation of motion is to apply
Eq. (\ref{wrongweigh}) but to then remove the contribution to
$E^c$ by orthogonally projecting to the constraint surface.  An exact
orthogonal projection is numerically expensive; a very accurate
approximate algorithm, which is exactly orthogonal but does not
quite complete the projection, was developed in \cite{Moore1}
(which is where the interested reader should go for more thorough
details).  It is also shown there that this technique gives a value
for $\Gamma_\mu$ equal to $\Gamma_d/2$, in accord with the fluctuation
dissipation relation, a good check.

In the case of Yang-Mills Higgs theory, it is no longer true that 
$E^c=0$; instead the Gauss constraints stipulate that
\begin{equation}
\forall x,a \, \quad -{\rm Re} \left( \Pi^{\dagger}(x) i \tau^a \Phi(x)
\right) + \sum_i \left[
E_i^a(x) - (U_i^{\dagger}(x-i)E_i(x-i)U_i(x-i))^a \right] = 0 \, .
\end{equation}

To analyze the application of a chemical potential it is convenient to
consider an orthonormal basis for the momenta $E,\Pi$ which
partitions into 4 subcatagories:
the $E^*$; the radial components of $\Pi$; an orthonormal 
basis of the linear combinations of
degrees of freedom which are forced zero by the Gauss constraints,
$P^c$; and an orthonormal basis of the remaining degrees of freedom,
$P^*$.  $P^c$ and $P^*$ are mixtures of the $E^c$ and Higgs
momentum degrees of freedom.  The definition of $\dot{N}_{CS}$ can
be written as
\begin{equation}
2 \pi^2 \dot{N}_{CS} = B \cdot E = B^* \cdot E^* + B^c \cdot P^*
+ B^c \cdot P^c \, ,
\end{equation}
where the dot products involving the $P$ are between $B$ and the $E$
field components of the $P$.  As above, applying a chemical
potential using this definition of $\dot{N}_{CS}$ will excite 
violations of the Gauss constraints; again the solution is to say
that the definition of $\dot{N}_{CS}$ is
\begin{equation}
2 \pi^2 \dot{N}_{CS} = B^* \cdot E^* + B^c \cdot P^*  \, ,
\end{equation}
which is equivalent on the constraint manifold and therefore does not
require a change in the program code which computes it.  All that
changes is that the momenta should be orthogonally projected
to the constraint manifold each update.  The algorithm for this 
orthogonal projection is the same as the algorithm proposed for use
in thermalizing Yang-Mills Higgs theory in \cite{Moore1} and used
in this paper for the canonical ensemble algorithm.

\subsection{Results for $N_{CS}$ motion in Yang-Mills Higgs theory}

\begin{table}
\begin{tabular}{|c|c|c|c|c|c|c|} \hline
Theory & $\beta_L$ & $m_{H0}$ & Phase & $t$ 
& $\kappa_d$ & ``expected errorbar'' \\ \hline \hline
YM & 6 & N.A. & N.A. & 1000 & $1.05 \pm .11$ & $\pm .18$ \\ \hline
YMH & $\simeq 8$ & -.3223 & symmetric 
& 4000 & $0.85 \pm .07$ & $\pm .14$ \\ \hline
YMH & $\simeq 8$ & -.3223 & broken    
& 8000 & $0.111 \pm .009$ & $\pm .037$ \\ \hline
YMH & $8$ & -.50 & deep broken & 
3000 & $0.059 \pm .008$ & $\pm .04$ \\ \hline
\end{tabular}
\caption{\label{NCStable} Diffusion rates of $N_{CS}$ in Yang-Mills
and Yang-Mills Higgs theory.  In the last column the bare Higgs 
mass squared was $m_{H0}=-0.5$ in lattice units, placing the system
deep in the broken phase.  The ``expected errorbar'' is the errorbar
which would occur if the diffusion were made out of Poisson
distributed integer steps.}
\end{table}

\begin{figure}
\centerline{\psfig{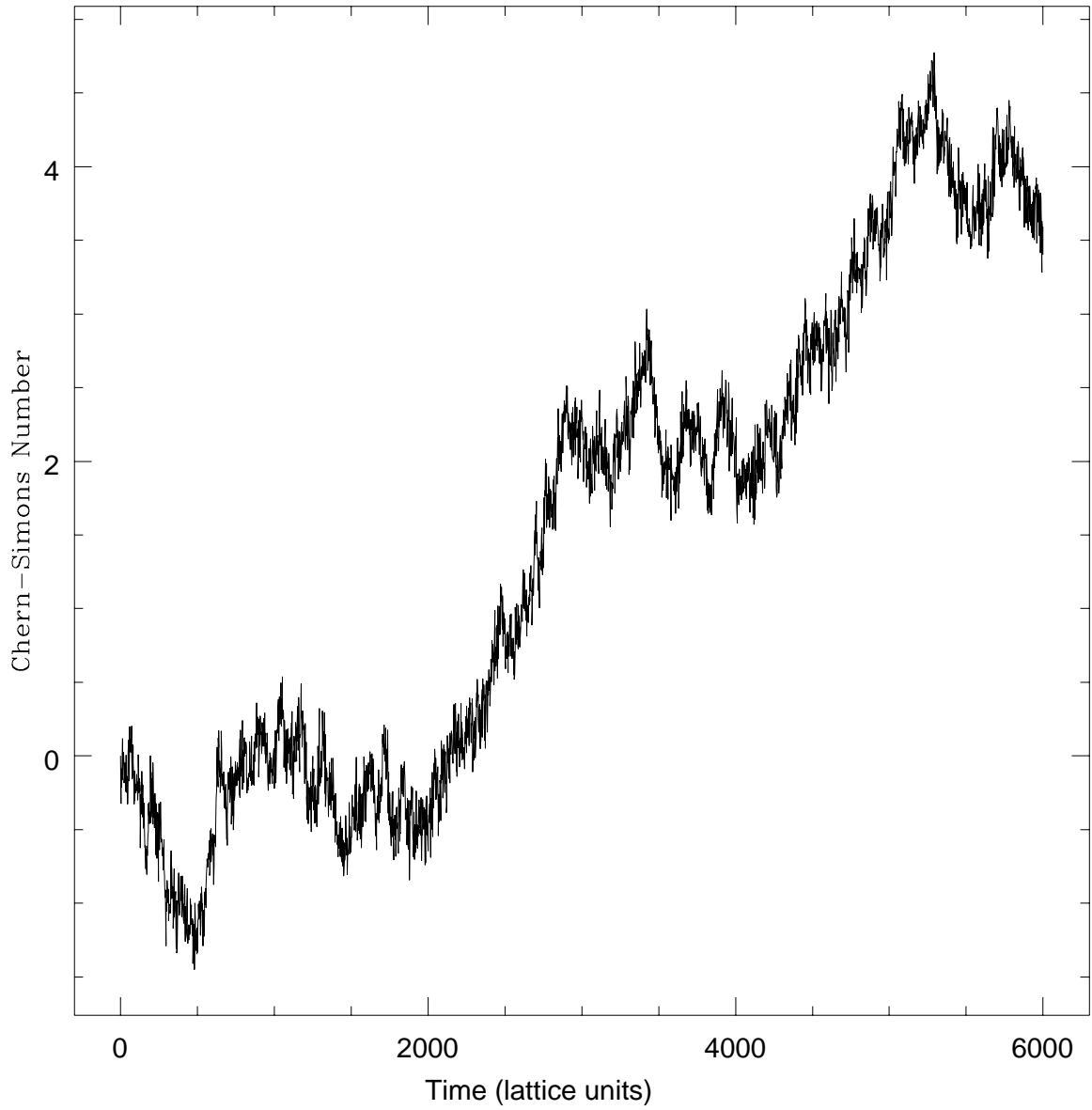}}
\caption{ \label{NCSdiffusion} Chern-Simons number diffusing in the
broken electroweak phase in a $16^3$ box at $\beta_L \simeq 8$.  The
diffusion does not resemble sudden discrete jumps between integer
spaced plateaus.}
\end{figure}

\begin{figure}
\centerline{\psfig{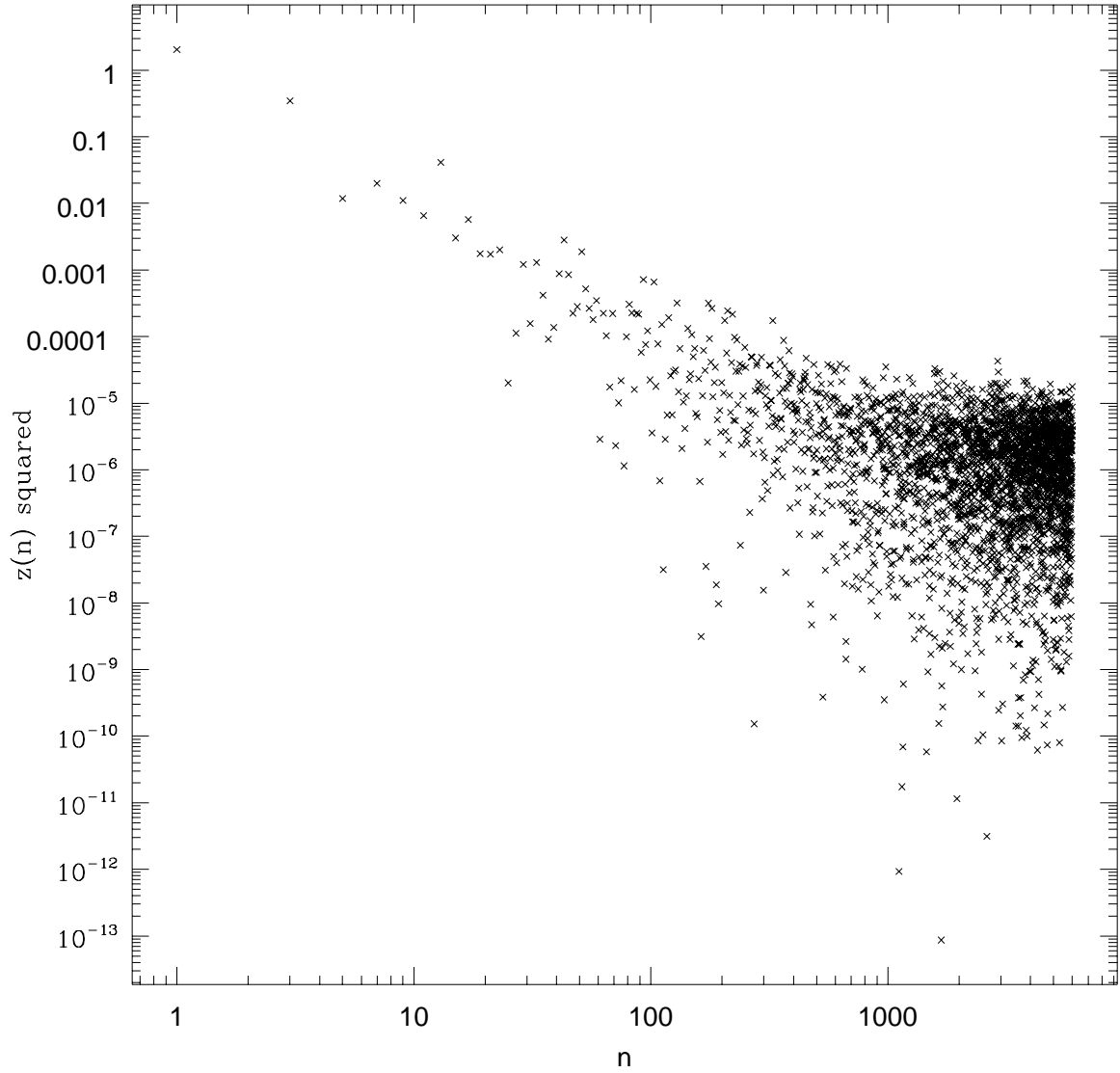}}
\caption{\label{NCSdiffusionft} Sine transform of the previous data.
The spectrum is in excellent agreement with white noise (constant
power) plus a Brownian signal (power $\propto n^{-2}$).  As discussed
in the text, the Brownian motion is probably a lattice artefact.}
\end{figure}

Now we will apply these tools to investigate the motion of $N_{CS}$ 
in Yang-Mills Higgs theory in each phase.  Data for the diffusion 
rate in Yang-Mills and Yang-Mills Higgs theory are given in Table 
\ref{NCStable}.  As has become standard in the literature, the rate
is expressed in terms of 
\begin{equation}
\kappa_d \equiv \Gamma_d (\pi \beta_L)^4 \; {\rm lattice \; units}
\quad {\rm or} \quad \Gamma_d (\alpha_w T)^{-4} \; {\rm physical
\; units} \, .
\end{equation}
All data are for $16^3$ lattices, which should be large enough to
achieve the infinite volume limit for the values of $\beta_L$ used
\cite{AmbKras}.  The Yang-Mills Higgs data are all for
(bare) $\lambda_L=0.20$.  To illustrate the data analysis technique,
Figure \ref{NCSdiffusion} shows a 
6000 lattice unit section of the evolution of the Yang-Mills Higgs 
system in the broken phase, and Figure \ref{NCSdiffusionft}
gives the coefficients of the sine transform, clearly showing the
$A/n^2 + B$ behavior as well as the wide scatter associated with
the log of the square of a Gaussian quantity.

There are three pieces of evidence which make it
difficult to believe that the observed diffusion constant in the
broken phase represents genuine infrared, topology changing physics.
The first is rather subtle; it involves error bars.  A series of
integer steps in $N_{CS}$ would not appear as perfect diffusive
motion, and this would be reflected in the error bars in
the extracted diffusion constant; if the steps
were Poisson distributed, as seems reasonable, then $N$ steps would
give error bars in $\Gamma_d$ of $\Gamma_d/\sqrt{N}$.  The number of
steps, based on $\Gamma_d$, is $\Gamma_d V t$, and the error estimate
based on this reasoning is shown in the last column of the table.  
In the broken phase the actual
error bars are much too small, suggesting that the diffusive process
is somehow much smoother than expected.  This would happen if it was
the accumulation of a large set of small, ultraviolet shifts in
$N_{CS}$.  

Ambjorn and Krasnitz have shown that, at leading order, 
the ultraviolet theory behaves like an abelian theory, and shifts in
$N_{CS}$ are elastic, and are restored a moment later
\cite{AmbKras}.  But they also
show that such shifts occur at a rate which diverges as $a \rightarrow 0$,
so if at next to leading order such shifts are not perfectly restored,
it could still contribute an effect which would not vanish as $a
\rightarrow 0$.  This motivates the possibility of a spurious signal which
remains constant in the small $a$ limit, which would neatly explain
the behavior of the diffusion rate observed above.

For a further piece of evidence, we applied a very large chemical 
potential, $\mu = 6/ \beta_L$, to a $16^3$ lattice of Yang-Mills Higgs
plasma in the broken phase just below the phase transition temperature.
Over a time period of $t=4000$, $N_{CS}$ changed by 11.4, which would
correspond to $\kappa_d=.088 \pm .026$ if $\dot{N}_{CS}$ rises linearly
with $\mu$ out to $6/\beta_L$.  Since this agrees with the 
measured value of $\kappa_d$, we conclude that the response
is indeed linear.  However, if
there were a substantial free energy barrier we would expect the rate
to rise with sinh$(2\beta_L \mu)$, which it clearly does not.
Since symmetry is broken strongly enough at the value of $\lambda_L$
we are considering that there should be a substantial free energy
barrier, this implies that the effect we observe is not the system
jumping that barrier, but some other process, presumably the ultraviolet
effect suggested above.  This evidence is in the same spirit as
the first piece.

For a final, strong piece of evidence, consider the last line in
the table.  Here we evolved a volume of Yang-Mills Higgs fluid with a
large negative Higgs mass squared, so it was very deep in the broken
phase; in fact we measured $\beta_L^2 \Phi^{\dagger} \Phi=98$,
over 5 times its value in
the broken phase just after the phase transition for the parameter
values we have concentrated on in most of this paper.  This corresponds,
in physical units, and after subtracting off the ultraviolet
$1/a$ contribution, to $\phi = 4.7 gT$.  The tree level Sphaleron energy
for this value of $\phi$ is $4.7 \times 4 \pi B(\lambda/g^2) T \simeq 
100T$ \cite{Manton}.  For such a large value, the semiclassical Sphaleron 
approximation should be very reasonable, and the diffusion rate should
have an immense exponential suppression, by a factor on order $\exp(-100)$
\cite{Manton,McLerran}; yet the measured rate is
only mildly lower than that barely inside the broken phase.  
This diffusion of $N_{CS}$ {\it must} be a lattice artefact, unless
our understanding of baryon number violation deep in the broken
phase is completely wrong.  The slight
decline in the rate relative to the rate ``just'' inside the broken phase 
probably comes about because $\phi$ is larger even than the
lattice inverse spacing, so even the ultraviolet lattice modes
have their populations mildly suppressed.

We should point out that, while the error bars for diffusion in Yang-Mills
theory and in the symmetric phase of Yang-Mills Higgs theory 
are also smaller than expected, they
are not as much smaller, and the argument that $N_{CS}$ should diffuse
in integer steps is invalid in the unbroken phase.  Further, it has been
demonstrated that the diffusion rate falls off for small volumes, which
shows that the dominant contribution is indeed infrared and should 
reflect real physics \cite{AmbKras}.  It is likely
that a small part $\sim 10 \%$
of the measured rate in the symmetric phase and in Yang-Mills theory
arises from ultraviolet artefacts; in fact the results here could be
viewed as a calibration of that contribution, since the ultraviolet 
behavior of lattice Yang-Mills Higgs theory should be about the same in
each phase.

\section{Bubble wall surface tension}

In this appendix we present the details of the calculation of the
bubble wall surface tension.  The determination has two parts:  first
we identify the bubble wall surfaces, then we show how to extract the
surface tension from the shape of the surface (averaging over a 
large sample of surfaces).  The reader can skip the first subsection 
and just take the wall surface position to be known if
they are uninterested in numerical details.

\subsection{Finding the surface}

Suppose we have the values of $\Phi$ and the connections $U$ at some point in 
time and we want to determine where the phase boundary is.  The
problem is that $\Phi$ contains a lot of ultraviolet fluctuations, so
a simple threshold definition of the two phases does not work.  We
will need to coarse grain or smooth the fields.  Define the once smoothed
Higgs field $\Phi_1$ as
\begin{equation}
\Phi_1(x) = \frac{1}{4} \Phi(x) + \frac{1}{8} \sum_i \Big[  U_i(x) \Phi(x+i)
+ U^{\dagger}_i(x-i) \Phi(x-i) \Big] \, .
\end{equation}
The sum just averages over nearest neighbors, parallel transporting along
the shortest path.  $\Phi_{n+1}$ is defined by applying the same
averaging process to $\Phi_n$.  For large $n$, $\Phi_n$ is approximately
$\Phi$ averaged over a Gaussian envelope of variance $3n/4$.  The averaging
very quickly removes ultraviolet fluctuations.  It also very slowly 
degrades the condensate, because the averaging includes an averaging 
over different parallel transport paths, and the trace of any Wilson
loop is less than ${\rm Tr} I$.  
After several smoothings (we use $\Phi_8$, but
for a finer lattice or a larger value of $\lambda_L$ we would need
to use more) the phases become much more distinct, although one still
cannot distinguish them by setting a threshold.

To work with a real $c$ number quantity we take $\Phi_n^{\dagger} \Phi_n$.
Label the three directions $x_1, x_2,$ and $x_3$, with $x_3$ the long 
direction in the lattice.  For each column (fixed $x_1$ and $x_2$) we
smooth along the column with a Gaussian envelope of width large enough 
to make the two phases distinguishable by a threshold, but smaller than
the typical wall thickness (which for our value of $\lambda_L$ and $\beta_L$
will turn out to be on order 15).  In practice we find a width $\simeq 7$
is sufficient.  (If it is impossible to smooth enough that the two
phases can always be distinguished by a threshold without making the
smoothing length larger than the wall thickness, then one should have
used more gauge invariant smoothings earlier.)  Call the function we
get by doing this smoothing $f(x_1,x_2,x_3)$.  The only problem left
is to choose a threshold by which to define the phases.  We do this
as follows.  First, for each column we find the minimum and maximum
values of $f$; average these over the columns; call them $f_{min}$
and $f_{max}$.  We assume that any point with $f$ more than halfway from
$f_{min}$ to $f_{max}$ is in the broken phase and any point less
than $0.2$ of the way is in the symmetric phase, and from this 
definition we find the average value of $f$ in each phase.  Then we set
a threshold some percentage of the way between the average symmetric
phase value of $f$ and the average broken phase value of $f$; at 
equilibrium the choice is $30 \%$, reflecting that fluctuations in $f$
are much larger in the broken phase; but for a heavily supercooled system
we set the threshold higher, $40 \%$, because the symmetric phase
fluctuations become larger.

With the threshold set, we define the wall height function $z(x_1,x_2)$
by starting at the value of $x_3$ in the $x_1,x_2$ 
column where $f$ is minimum and
incrementing $x_3$ until the first point is reached where the threshold
is exceeded; $z(x_1,x_2)$ is chosen as the value of 
$x_3$ where this happens.  The other surface is found by moving backward.  

The definition of the surface described here is imperfect; it assumes that
the surface is single valued (never bending more than $90^\circ$ from
horizontal) and it smooths the surface at a length scale on order
the Gaussian envelope radius used to define the smoothed Higgs field
$\Phi_n$.  However, the surface is not really well defined on 
length scales much shorter than the wall thickness, and we will be
most interested in infrared scales on which the surface should look
relatively flat, so neither problem will be of consequence.

\subsection{Determining the surface tension}

Take the height function $z(x_1,x_2)$ to be known; how can one determine
the surface tension from it?

The free energy of a surface with surface tension $\sigma$ and height
function $z$ is the area of the surface times the surface tension,
\begin{equation}
F = \sigma \int d^2 x \sqrt{ 1 + (\nabla z)^2 } \, .
\end{equation}
At finite temperature, surface waves will be excited and the partition
function for the appearance of the surface will be
\begin{equation}
Z = \int {\cal D} z \exp ( - \beta F ) \, ,
\end{equation}
which is the partition function of a simple two dimensional Euclidean
field theory.  In the infrared we can expand $F$ as
\begin{equation}
F = \sigma \int d^2 x \left( 1 + \frac{1}{2} (\nabla z)^2 - \frac{1}{8}
( (\nabla z)^2 )^2 + \ldots \right) \, .
\end{equation}
The first term contributes a $z$ independent constant and the remaining
terms describe a free massless field theory plus nonrenormalizeable
operators.  These operators will cause some renormalization of $\sigma$
between an ultraviolet scale and the deep infrared, but because they
are nonrenormalizeable $\sigma$ will have an unambiguous deep infrared
limit, which is the value of interest in any calculation where the
surface tension is a useful concept.  concentrating on the infrared
and remembering that there may be corrections for ultraviolet modes,
and taking the area over which the surface is stretched to be
an $L \times L$ square, 
\begin{equation}
\beta F \simeq \frac{\sigma}{T} \int_0^L \int_0^L d^2x \left( 
1 + \frac{1}{2} (\nabla z)^2 \right) \, .
\end{equation}
This free field theory is trivially solved by Fourier transformation.
Defining
\begin{equation}
\tilde{z}_{\vec{n}} = \int \frac{d^2x}{L^2} z(x) e^{2 \pi i \vec{n}
\cdot x / L}
\end{equation}
we find
\begin{equation}
\beta F = \sum_{\vec{n}} \frac{4 \pi^2 \sigma n^2 \tilde{z}_n^2}{2T}
\end{equation}
and by equipartition
\begin{equation}
\langle \tilde{z}^2_n \rangle = \frac{T}{4 \pi^2 \sigma n^2} \, .
\label{powerspectrum}
\end{equation}

This expression should be obeyed in the limit of large $L/n$; for 
smaller values of $L/n$ it will have corrections 
because of the nonrenormalizeable
operators, and in the numerical calculation it will have corrections
in the ultraviolet because of the properties of the numerically determined
bubble wall surface mentioned earlier.  To compute the surface tension
it will then be necessary not only to average the $\tilde{z}_n$ over
a large sample of walls, but to perform an extrapolation to zero $n$.
Because the corrections to the free field theory are
nonrenormalizeable, the corrections to Eq. (\ref{powerspectrum}) will 
appear as powers of $(n/L)^2$; there will be no logarithmic corrections.
As discussed in the text we performed the large $L/n$ extrapolation 
by fitting $n^2 \langle
\tilde{z}_n^2 \rangle$ to $A\exp(-B n^2)$.  We do not know of a good
physical motivation for this form except that it resembles what a
Gaussian smoothing of the wall surface would do to the $\tilde{z}_n$.
In practice we could equally well fit only the first few points to a
straight line; the result is about the same.

\pagebreak

\begin{thebibliography}{99}
\bibitem{Sakharov} A. Sakharov, JETP Lett. {\bf 6} (1967) 24.
\bibitem{tHooft} G. t'Hooft, Phys. Rev. Lett. {\bf 37} (1976) 8.
\bibitem{KRS} V. Kuzmin, V. Rubakov, and M. Shaposhnikov, Phys. 
	Lett. {\bf D 30} (1985) 36. 
\bibitem{McLerran} P. Arnold and L. McLerran, Phys. Rev. {\bf D 36}
        (1987) 581.
\bibitem{TurokZadrozny} N. Turok and J. Zadrozny, Phys. Rev. Lett. 
	{\bf 65} (1990) 2331; Nucl. Phys. {\bf B 358} (1991) 471. 
\bibitem{MSTV} L. McLerran, M. Shaposhnikov, N. Turok, and M. Voloshin, 
	Phys. Lett. {\bf B 256} (1991) 451.
\bibitem{CKN} A. Nelson, D. Kaplan, and A. Cohen, Nucl. Phys. {\bf B 373}
	(1992) 453.
\bibitem{DHSS} M. Dine, P. Huet, R. Singleton and L. Susskind, 
	Phys. Lett. {\bf B257}, 351 (1991).
\bibitem{KLRS} K. Kajantie, M. Laine, K. Rummukainen, and M.
	Shaposhnikov, Nucl. Phys. {\bf B 458} (1996) 90.
\bibitem{MSSMguys} J. Cline and K. Kainulainen, CERN-TH/96-76,
	hep-ph/9605235; M. Losada, RU-96-25, hep-ph/9605266;
	M. Laine, Nucl. Phys. {\bf B 481}, 43 (1996).
\bibitem{DolanJackiw} L. Dolan and R. Jackiw, Phys. Rev. {\bf D 9}
	(1974) 3320.
\bibitem{Espinosa} P. Arnold and O. Espinosa, Phys. Rev. {\bf D 47}
	(1993) 3546; Phys. Rev. {\bf D 50} (1994) 6662.
\bibitem{Fodor} Z. Fodor and A. Hebecker, Nucl. Phys. {\bf B 432}
	(1994) 127.
\bibitem{FKRS1} K. Farakos, K. Kajantie, K. Rummukainen, and M.
	Shaposhnikov, Nucl. Phys. {\bf B 425} (1994) 67.
\bibitem{FKRS} K. Farakos, K. Kajantie, K. Rummukainen, 
	and M. Shaposhnikov, Nucl. Phys. {\bf B 442}
	(1995) 317.
\bibitem{KLRSresults} K. Kajantie, M. Laine, K. Rummukainen, and M.
	Shaposhnikov, Nucl. Phys. {\bf B 466} (1996) 189.
\bibitem{GrigRub} D. Grigorev and V. Rubakov, Nucl. Phys. {\bf B 299}
	(1988) 248.
\bibitem{Amb1} J. Ambjorn, M. Laursen, and M. Shaposhnikov, Nucl. Phys.
	{\bf B 316} (1989) , 483.
\bibitem{Ambjornetal} J. Ambjorn, T. Askgaard, H. Porter, and M.
        Shaposhnikov, Nucl. Phys. {\bf B 353} (1991) 346.
\bibitem{AmbKras} J. Ambjorn and A. Krasnitz, Phys. Lett. {\bf B 362}
	(1995) 97.
\bibitem{Moore1} G. D. Moore, Nucl. Phys. {\bf B 480} (1996) 657.
\bibitem{Moore2} G. D. Moore,  Nucl. Phys. {\bf B 480} (1996) 689.
\bibitem{newguys} W. Tang and J. Smit, Nucl. Phys. {\bf B 482} (1996) 265.
\bibitem{early} K. Kajantie, K. Rummukainen, and M. Shaposhnikov,
	Nucl. Phys. {\bf B 407} (1993) 356;
	K. Farakos, K. Kajantie, K. Rummukainen, 
	and M. Shaposhnikov, Nucl. Phys. {\bf B 425} (1994) 67; 
	Phys. Lett. {\bf B 336} (1994) 494.
\bibitem{Landsmann} N. Landsman, Nucl. Phys. {\bf B 332} (1989) 498.
\bibitem{Bodekernew} D. Bodeker, AD-THEP-96-27, hep-th/9609170.
\bibitem{evans} T.S. Evans and D.A. Steer, Nucl. Phys. {\bf B 474} (1996)
	481.
\bibitem{BraatenI} E. Braaten and R. Pisarski, Nucl. Phys. {\bf B 337}
	(1990) 569.
\bibitem{Smilga} D. Bodeker, L. McLerran, and A. Smilga, Phys.
	Rev. {\bf D 52} (1995) 4675.
\bibitem{BraatenII} E. Braaten and R. Pisarski, Phys. Rev. {\bf D 42}
	(1990) 2156; Phys. Rev. {\bf D 46} (1992) 1829.
\bibitem{Kogut} J. Kogut and L. Susskind, 
	Phys. Rev. {\bf D 11} (1975) 395.
\bibitem{otherguys} F. Csikor, Z. Fodor, J. Hein, and J. Heitger,
	Phys. Lett. {\bf B 357} (1995) 156;
	F. Csikor, Z. Fodor, J. Hein, A. Jaster, and
	I. Montvay, Nucl. Phys. {\bf B 474} (1996) 421;
	M. Guertler, E. Ilgenfritz, J. Kripfganz, H. Perlt, and A. Schiller,
	hep-lat/9512022; hep-lat/9605042.
\bibitem{Kripfganz} J. Kripfganz, A. Laser, and M. Schmidt, HD-THEP-95-53,
	hep-ph/9512340.
\bibitem{Turok} N. Turok, Phys. Rev. Lett. {\bf 68} (1992) 1803.
\bibitem{Ignatius} K. Enqvist, J. Ignatius, K. Kajantie, and K.
	Rummukainen, Phys. Rev. {\bf D 45} (1992) 34415;
	J. Ignatius, K. Kajantie, H. Kurki-Suonio and M. Laine,
	Phys. Rev. {\bf D 49} (1994) 3854.
\bibitem{Dine} M. Dine, R. Leigh, P. Huet, A. Linde, and D. Linde, 
	Phys. Rev. {\bf D 46} (1992) 550.
\bibitem{Liu} B. Liu, L. McLerran, and N. Turok, Phys. Rev. 
	{\bf D 46} (1992) 2668.
\bibitem{Khlebnikov} S. Khlebnikov, Phys. Rev. {\bf D 46} (1992) 3226.
\bibitem{Arnold} P. Arnold, Phys. Rev. {\bf D 48} (1993) 1539.
\bibitem{Heckler} A. Heckler, Phys. Rev. {\bf D 51} (1995) 405.
\bibitem{MP1} G. Moore and T. Prokopec, Phys. Rev. Lett. {\bf 75}
	(1995) 777.
\bibitem{MP2} G. Moore and T. Prokopec, Phys. Rev. {\bf D 52}
	(1995) 7182.
\bibitem{Laine} H. Kurki-Suonio and M. Laine, Phys. Rev. {\bf D 54}
	(1996) 7163.
\bibitem{Jeon} S. Jeon, Phys. Rev. {\bf D 52} (1995) 3591.
\bibitem{Krasnitz} A. Krasnitz, Nucl. Phys. {\bf B 455} (1995) 320.
\bibitem{Krasprivate} A. Krasnitz, private communication.
\bibitem{NTrev} N. Turok, in ``Perspectives on Higgs Physics'', 
	Ed. G. Kane, World Scientific, 1993.
\bibitem{GST}  D. Grigoriev, M. Shaposhnikov and N. Turok,
	Phys. Lett {\bf 275B} (1992) 395.
\bibitem{DineThomas} M. Dine and S. Thomas, Phys. Lett. {\bf B 328}
	(1994) 73.
\bibitem{JPT} M. Joyce, T. Prokopec and N. Turok, 
	Physical Review {\bf D53}, 2930 (1996); 
	Physical Review {\bf D53}, 2958 (1996); 
	Physical Review Letters {\bf 75}, 1695 (1995).
\bibitem{Dinearg} See e.g. M. Dine, Nucl.Phys.Proc.Suppl.
	{\bf 37A}, 127 (1994).
\bibitem{KhlebShap} S. Khlebnikov and M. Shaposhnikov, Nucl. Phys.
	{\bf B308} (1988) 885.
\bibitem{RubShap}  V. Rubakov and M. Shaposhnikov, CERN-TH-96-13,
	hep-ph/9603208.
\bibitem{Manton} F. Klinkhamer and N. Manton, Phys. Rev. {\bf D 30}
	(1984) 2212.

\end{thebibliography}
\end{document}